\documentclass{aa}  
\pdfoutput=1
\usepackage[varg]{txfonts} 
\usepackage{natbib}
\usepackage{float}
\usepackage{graphicx}
\usepackage{amsmath}	
\usepackage{mathtools}
\usepackage{amssymb}	
\usepackage{hyperref}
\hypersetup{
    colorlinks   = true,
    urlcolor     = blue,
    linkcolor    = blue,
    citecolor   = blue
}
\usepackage{multirow}
\usepackage{tabularx}
\usepackage[group-separator={,}]{siunitx}
\usepackage{bm}
\usepackage{xcolor}
\usepackage[export]{adjustbox}
\usepackage{url}
\usepackage[switch]{lineno}
\usepackage[rightcaption]{sidecap}

\begin{document} 

   \title{KiDS-Legacy: WIMP dark matter constraints from the cross-correlation of weak lensing and Fermi-LAT gamma rays}
   \titlerunning{Cross-correlation of weak lensing and gamma rays}
   \authorrunning{S. Zhang et al.}

    \author{Shiyang Zhang\inst{1}
        \and
        Hendrik Hildebrandt\inst{1}
        \and
        Ziang Yan\inst{1,2}
        \and
        Tilman Tr\"{o}ster\inst{3} %?
        \and
        Athithya Aravinthan\inst{4}
        \and
        Marika Asgari\inst{5}
        \and
        Deaglan J. Bartlett\inst{6, 7}
        \and
        Maciej Bilicki\inst{8}
        \and
        Dominik Els\"{a}sser\inst{9}
        \and
        Catherine Heymans\inst{1, 10}
        \and
        Benjamin Joachimi\inst{11}
        \and
        Lauro Moscardini\inst{12,13,14}
        \and
        Dennis Neumann\inst{15} %?
        \and
        Anya Paopiamsap \inst{16}
        \and
        Robert Reischke \inst{1, 17}
        \and
        Benjamin St\"{o}lzner \inst{1}
        }

\institute{Ruhr University Bochum, Faculty of Physics and Astronomy, Astronomical Institute (AIRUB), German Centre for Cosmological Lensing, 44780 Bochum, Germany\\ 
\email{shiyang@astro.ruhr-uni-bochum.de}
\and
Graduate School of Science, Nagoya University, Furocho, Chikusa-ku, Nagoya, Aichi, 464-8602, Japan 
\and
Institute for Particle Physics and Astrophysics, ETH Z\"{u}rich, Wolfgang-Pauli-Strasse 27, 8093 Z\"{u}rich, Switzerland % Tilman #3
\and
RAPP Center and Theoretische Physik IV, Fakult\"{a}t f\"{u}r Physik \& Astronomie, Ruhr-Universit\"{a}t Bochum, 44780 Bochum, Germany % # Athy #4
\and
School of Mathematics, Statistics and Physics, Newcastle University, Herschel Building, NE1 7RU, Newcastle-upon-Tyne, UK % Marika #5
\and
Astrophysics, University of Oxford, Denys Wilkinson Building, Keble Road, Oxford, OX1 3RH, UK % Deaglan #6
\and
CNRS \& Sorbonne Universit\'{e}, Institut d’Astrophysique de Paris (IAP), UMR 7095, 98 bis bd Arago, F-75014 Paris, France % Deaglan #7
\and
Center for Theoretical Physics, Polish Academy of Sciences, al. Lotników 32/46, 02-668 Warsaw, Poland % Maciej #8
\and
Department of Physics, TU Dortmund University, D-44221 Dortmund, Germany % Dominik #9
\and
Institute for Astronomy, University of Edinburgh, Royal Observatory, Blackford Hill, Edinburgh, EH9 3HJ, UK. % Cathrine #10
\and
Department of Physics and Astronomy, University College London, Gower Street, London WC1E 6BT, UK % Benjamin J #11
\and
Dipartimento di Fisica e Astronomia "Augusto Righi" - Alma Mater Studiorum Università di Bologna, via Piero Gobetti 93/2, I-40129 Bologna, Italy % Lauro #12
\and
Istituto Nazionale di Astrofisica (INAF) - Osservatorio di Astrofisica e Scienza dello Spazio (OAS), via Piero Gobetti 93/3, I-40129 Bologna, Italy  % Lauro #13
\and
Istituto Nazionale di Fisica Nucleare (INFN) - Sezione di Bologna, viale Berti Pichat 6/2, I-40127 Bologna, Italy % Lauro #14
\and
Leiden Observatory, Leiden University, Einsteinweg 55, 2333 CC Leiden, The Netherlands % Dennis #15 
\and
Institut de Ciències del Cosmos, Universitat de Barcelona (ICCUB),  Martí i Franquès, 1, 08028 Barcelona, Spain  % Anya #16
\and
Argelander-Institut für Astronomie, Universität Bonn, Auf dem Hügel 71, D-53121 Bonn, Germany % Robert #17
}

   \date{Received XXX, XXXX; accepted YYY, YYYY}
 
  \abstract
  {Dark matter dominates the matter content of the Universe, and its properties can be constrained through large-scale structure probes such as the cross-correlation between the unresolved gamma-ray background (UGRB) and weak gravitational lensing.
  We analysed 15 years of Fermi–LAT data, constructing UGRB intensity maps in ten energy bins (0.5–1000 GeV), and cross-correlate them with KiDS-Legacy shear in six tomographic bins.
  The measurements were performed using angular power spectra estimated with the pseudo-$C_\ell$ method. No significant cross-correlation is found.
  Based on this non-detection, we present 95\% upper bounds on the weakly interacting massive particle (WIMP) decay rate $\Gamma_{\rm dec}$ and velocity-averaged annihilation cross-section $\langle\sigma_{\rm ann} v\rangle$ as functions of mass. We compare our results with bounds from other cosmological tracers and from local probes, and found them to be complementary, particularly at low masses ($\rm GeV/TeV$). In addition, using a \textit{Euclid}-like lensing survey cross-correlated with Fermi–LAT, we forecast $\sim$2 times tighter limits, highlighting the potential of forthcoming data to strengthen constraints on dark matter annihilation and decay.}

   \maketitle

\section{Introduction}
\label{Sec:Intro}

Dark matter (DM) constitutes a significant portion of the Universe, amounting to approximately 27\% of the total mass-energy content according to constraints from \textit{Planck} \citep{2020A&A...641A...6P}. 
Unlike baryonic matter, DM appears invisible with its presence inferred from its gravitational effects on visible matter and the large scale structure of the Universe. The first evidence for DM came from Fritz Zwicky’s 1930s study of the Coma Cluster \citep{zwicky1979masses}, followed by the discovery of flat galaxy rotation curves \citep{rubin1970rotation}. 
Although precise measurements of the DM cosmological abundance and statistical properties of its distribution in the Universe have been achieved, its microscopic properties, such as its non-gravitational interactions and composition, are still mostly unknown.

One of the most studied candidates for DM is the weakly interacting massive particle (WIMP). WIMPs are hypothetical particles that interact only through the weak nuclear force and gravity \citep{bertone2005particle}. With masses ranging from approximately 1 GeV to 1 TeV, thermally produced WIMPs in the early Universe would naturally exhibit a relic density that matches the observed DM abundance. This is due to their decoupling from the primordial plasma in the non-relativistic regime and their weak interactions with lighter standard model particles (\citealt{lee1977cosmological}; \citealt{gunn1978some}).

The weak coupling of WIMPs allows us to test the hypothesis of WIMP DM and offers a promising method to indirectly probe it. Assuming WIMPs are the building blocks of large-scale structure in the Universe, it is likely that WIMPs within DM haloes may annihilate or decay into detectable standard model particles. These particles, known as final state particles, can then interact to produce gamma-rays, neutrinos, and cosmic rays. Searching for the primary and secondary radiation of standard model particles from the DM annihilation process is known as the indirect detection of DM (see e.g., \citealt{cirelli2012indirect} for a review).

Various instruments have been utilised in efforts to detect DM annihilation products. These include Cherenkov telescopes such as the High Energy Stereoscopic System (H.E.S.S., e.g. \citealp{abramowski2011search, abramowski2012search, collaboration2018search}), Very Energetic Radiation Imaging Telescope
Array System (VERITAS, e.g. \citealp{archambault2017dark}) and the Florian Goebel Major Atmospheric Gamma-ray Imaging Cherenkov (MAGIC) telescopes (e.g. \citealp{aleksic2011searches, ahnen2016limits, acciari2022combined, abe2023search}); neutrino telescopes such as IceCube \citep{abbasi2023search}; the High Altitude Water Cherenkov Experiment (HAWC, e.g. \citealp{aartsen2018search, albert2018dark}); and the Fermi Large Area Telescope (Fermi-LAT, e.g. \citealp{ackermann2010constraints, ahnen2016limits, mcdaniel2024legacy, totani2025JCAP...11..080T}). These instruments study the DM annihilation process by targeting regions predicted to have high DM densities in the local Universe, including the Galactic Centre (GC), and the Galactic halo and subhaloes where dwarf spheroidal galaxies (dSphs) of the Milky Way reside. Among the DM annihilation products, gamma-rays are the most investigated because they travel without being deflected by magnetic fields, allowing them to be traced back to their source and providing crucial insights into the spatial distribution of DM. At high energies, however, gamma-rays can be partially attenuated during propagation, so their observed flux carries integrated information about the intervening cosmic medium.

The unresolved gamma-ray background (UGRB) is a significant focus in DM studies on larger scales, offering a potential probe for the indirect detection of DM across cosmological distances (e.g., \citealt{fornasa2015nature}). 
The UGRB emission due to DM annihilation is expected to be anisotropic, arising from the underlying large–scale distribution of DM haloes and subhaloes (see e.g., \citealt{ando2006anisotropy}). 
In addition to potential DM signals, the UGRB comprises cumulative emissions from all unresolved astrophysical sources, including star-forming galaxies (SFGs, \citealt{ajello2020gamma, roth2021diffuse}), misaligned active galactic nuclei (mAGNs, \citealt{di2013diffuse}), blazars \citep{korsmeier2022flat}, millisecond pulsars \citep{calore2014diffuse}, and the interaction of cosmic rays with extragalactic background light (EBL) \citep{kalashev2009ultrahigh}. 

Weak gravitational lensing is a powerful tool for cosmological studies (see, e.g., \citealt{bartelmann2001weak}). It occurs when the gravitational field of the intervening large-scale structure induces slight distortions in the light paths of distant background objects. Weak lensing can serve as an unbiased tracer of the matter distribution, making it essential for studying the distribution and properties of DM \citep{camera2013novel}. Weak gravitational lensing can be measured statistically from the apparent correlations between the ellipticities of distant galaxies. By comparing these correlations to model predictions, one can constrain cosmological parameters such as $\Omega_{\rm m}$ and $\sigma_8$ (see, e.g., \citealt{Secco2022PhRvD.105b3515S, amon2022dark, dalal2023hyper, wright2025arXiv250319441W, stolzner2025arXiv250319442S}). 

In recent years, cross-correlations between UGRB maps and weak lensing data have become a powerful tool for probing DM properties. \citet{shirasaki2016cosmological} performed a cross-correlation analysis using 7 years of Fermi-LAT gamma-ray data with weak lensing surveys from the Canada-France-Hawaii Telescope Lensing Survey (CFHTLenS) and the Red Cluster Sequence Lensing Survey (RCSLenS). Building on this, \citet{troster2017cross} cross-correlated the UGRB maps from 8 years of Fermi-LAT data with tomographic weak lensing data from CFHTLenS, RCSLenS, and the Kilo Degree Survey (KiDS-450). The work from \citet{shirasaki2018correlation} applied a similar approach using the first-year data from the Hyper Suprime-Cam Subaru Strategic Programme (HSC-SSP). These works are all consistent with non detections and demonstrate that weak lensing data can provide significant information for constraining DM properties. \citet{ammazzalorso2020detection}, for the first time, achieved a $5.3 \sigma$ detection of a cross-correlation signal between gamma-rays from 9 years of Fermi-LAT UGRB and tomographic weak lensing data from Dark Energy Survey Y1 (DES Y1). More recently, \citet{thakore2025high} detected an $8.9 \sigma$ cross-correlation using 12 years of Fermi-LAT UGRB and DES Y3 lensing, dominated by large angular scales and consistent with blazar emission. While they also explored a DM component, they noted its degeneracy with astrophysical modelling and the tension with external gamma-ray limits, and therefore did not interpret it as evidence for DM. We further investigate the potential reasons for the discrepancy between the DES and KiDS weak-lensing cross-correlation results in Appendix~\ref{Sec:compare}.

In addition to weak lensing, other large-scale structure-based indirect detection approaches also provide valuable insights into DM properties. Spatial distributions of galaxies offer a high signal-to-noise ratio (S/N) for tracing matter fluctuations (\citealp{ando2014power, cuoco2015dark, regis2015particle, shirasaki2015cross, cuoco2017tomographic, ammazzalorso2018characterizing, bartlett2022constraints, kostic2023arXiv230410301K,paopiamsap2023constraints}), and their cross-correlations with gamma-rays yield significant signals (e.g. \citealp{paopiamsap2023constraints}). Additionally, cross-correlation studies with galaxy clusters have great potential, as these massive structures contain large amounts of DM, and their cross-correlation with gamma-rays has also been explored as a means of detecting DM signatures (\citealp{branchini2017cross, hashimoto2019measurement, colavincenzo2020searching, di2023constraining}). Other tracers including anisotropies of the Cosmic Microwave Background (CMB, \citealp{tan2020searching}),  CMB lensing \citep{fornengo2015evidence}, the Cosmic Infrared Background (CIB, \citealp{feng2017planck}), the late-time 21cm signal \citep{pinetti2020synergies}, the thermal Sunyaev-Zel’dovich effect \citep{shirasaki2020cross}, and high-energy neutrinos \citep{negro2023cross} are also used in cross-correlation studies.

Based on previous work, our study investigates the nature of DM by cross-correlating tomographic weak lensing data from the fifth data release of the Kilo Degree Survey (KiDS-Legacy, \citealt{Wright2024A&A...686A.170W}) with the energy-binned intensity map of the UGRB from 15 years of Fermi-LAT observations \citep{atwood2009large}. Properly modelling the contribution of astrophysical sources allows the UGRB measurements to better constrain the DM component (see, e.g., \citealt{fornasa2015nature} for a review). 

This paper is structured as follows: theoretical models are introduced in Sect.~\ref{Sec:Models}; we describe the data used for our analysis in Sect.~\ref{Sec:Data}; the data analysis methods and estimators are presented in Sect.~\ref{Sec:Methods}; the results are discussed in Sect.~\ref{Sec:Results}; we summarise our conclusions in Sect.~\ref{Sec:Conclusions}. We employ a flat $\Lambda$CDM cosmological model, with parameters taken from \citet{2020A&A...641A...6P}: $\left(h, \Omega_{\mathrm{DM}} h^2, \Omega_{\mathrm{b}} h^2, \sigma_8, n_{\mathrm{s}}\right) = \left(0.677, 0.119, 0.022, 0.810, 0.967\right) $.

\section{Theoretical models}
\label{Sec:Models}

Both observations, the weak gravitational lensing convergence (denoted as $\kappa$) and the gamma-ray emission intensity (denoted as $g$), are formalised as projections of three-dimensional fields. The general form of the projection is given by:
\begin{gather}
\label{Eq: proj}
u(\hat\theta) = \int \mathrm{d}\chi W^{u}(\chi) \delta(\chi\hat \theta, \chi), 
\end{gather}
where $u$ represents the projected field in the angular direction $\hat{\theta}$; $\chi$ is the radial comoving distance; $\delta(\chi\hat \theta, \chi)$ denotes the 3D fluctuation of the field $u$ at the coordinate $ (\chi\hat \theta, \chi)$;  $W^{u}(\chi)$ is the radial kernel that reflects the distribution of the field along the line-of-sight.

The theoretical model for the cross-correlation power spectrum between two such fields, $u$ and $v$, can be well estimated by the Limber approximation (\citealt{limber1953analysis, kaiser1992weak, kilbinger2017MNRAS.472.2126K}) when the radial kernels are significantly wider than their correlation length and at small scales ($\ell \gtrsim10$). It can be expressed as:
\begin{gather}
\label{Eq: Cell_general}
C_\ell^{uv} = \int\frac{\mathrm{d}\chi}{\chi^2}W^u(\chi)W^v(\chi)P_{uv}\left(k=\frac{\ell+1/2}{\chi(z)}, z(\chi)\right),
\end{gather}
we assume a spatially flat $\Lambda$CDM cosmology, where $z$ is the redshift, and $P_{uv}(k_\ell, z)$ is the cross power spectrum of the associated 3D fields $u$ and $v$.  It can be described as follows, with $k$ and $\ell$ being the modulus of the wave number and the angular multipole, respectively: 
\begin{gather}
\label{Eq: delta_delta}
\left\langle 
\delta_u(\boldsymbol{k})
\, \delta_v(\boldsymbol{k}')
\right\rangle
= (2\pi)^3 
\, \delta^{3}(\boldsymbol{k}+\boldsymbol{k}')
\, P_{uv}(k).
\end{gather}
where $\left<...\right>$ indicates the ensemble average of the quantity inside the brackets. Therefore, the cross-correlation angular power spectrum for lensing convergence $\kappa$ and gamma-ray intensity g can be expressed as:
\begin{gather}
\label{Eq: cell_cross}
C_\ell^{\rm g\kappa} = \int_{E_{\rm min}}^{E_{\rm max}}\mathrm{d}E \int^{\infty}_0 \mathrm{d}z  \frac{c}{H(z)} \frac{1}{\chi(z)^2} W_{\rm g}(E,z) W_{\kappa}(z)P_{\rm g\kappa}(k, z),
\end{gather}
where $E$ represents the gamma-ray energy; $E_{\rm min}$ and $E_{\rm max}$ define the upper and lower limits of the energy bin being integrated; $H(z)$ is the Hubble rate at redshift $z$; $W_{\rm g} (E,z)$ and $W_{\kappa}(z)$ are the window functions for gamma-ray intensity and gravitational lensing, respectively; and $P_{\rm g\kappa}(k, z)$ denotes the 3D cross power spectrum. 

\subsection{Gravitational lensing}
\label{Sec:gl}
\begin{figure}
   \centering
   \includegraphics[width=\hsize]{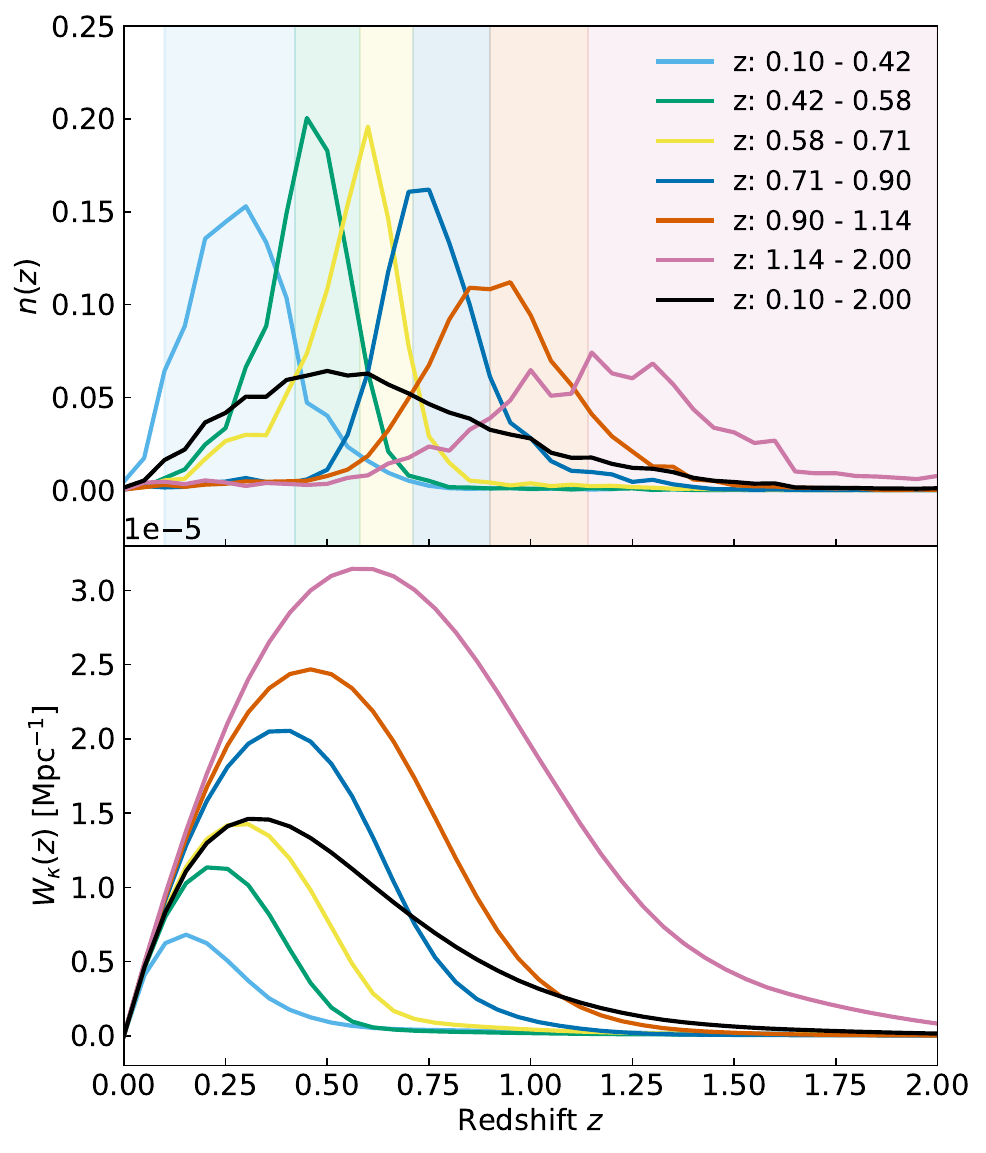}
   \caption{\textit{Top:} redshift distributions of the KiDS-Legacy gold-selected sample for tomographic bins, where the photometric redshift bin with $z$ in $[0.1, 2.0]$ is the sum weighted by the effective number densities listed in Table \ref{Tab:shape_noise}. The shaded vertical bands indicate the boundaries of the tomographic photometric redshift bins. \textit{Bottom:} weak lensing window functions for six tomographic bins in KiDS-Legacy.}
    \label{Fig: lensing_window}
    \end{figure}

The window function for gravitational lensing is given by (see, e.g. \citealp{bartelmann2010gravitational})
\begin{gather}
\label{Eq: w_gl}
W_\kappa(z) = \frac{3H_0^2\Omega_\mathrm{m}}{2} (1+z)~\chi(z)\int_z^\infty \mathrm{d}z^{\prime} \frac{\chi(z^{\prime}) - \chi(z)}{\chi(z^\prime)}n(z^\prime), 
\end{gather}
where $H_0$ is the Hubble constant;  $\Omega_{\rm m}$ represents the matter density parameter of the Universe; $n(z)$ is the redshift distribution of the source galaxies in each tomographic bin. The redshift distributions of the KiDS-Legacy gold-selected sample are shown in the top panel of Fig.~\ref{Fig: lensing_window} for the six bins used in this analysis. The integral term quantifies the lensing efficiency as a function of the source galaxies relative to the lens. The window functions for the KiDS-Legacy survey with six redshift bins are illustrated in the lower panel of Fig. \ref{Fig: lensing_window}.

\subsection{Gamma-rays from DM}
\label{Sec:dm}

DM particles, such as WIMPS, may annihilate upon collision, creating standard model particles and high-energy photons. 
We assume that DM is its own antiparticle (self-conjugate),  which we take as a standard benchmark in the literature, and the annihilation window function is given by (\citealp{ando2006anisotropy, fornengo2014particle,troster2017cross}):

\begin{align}
\label{Eq: w_ann}
W_{\rm g_{\rm ann}}(z, E) &= \frac{(\Omega_{\rm DM}\rho_{\rm c})^2}{4\pi}\frac{\left<\sigma_{\rm ann}v\right>}{2m_{\rm DM}^2}(1+z)^3  \Delta^2(z) \notag \\
&\quad \times\frac{\mathrm{d}N_{\rm ann}}{\mathrm{d}E}[E(1+z)]\rm e^{-\tau[\it z, E(1+z)]},
\end{align}
where $\Omega_{\rm DM}\rho_{\rm c}$ is the DM density in the Universe; $\rho_{\rm c}$ is the critical density of the Universe; $m_{\rm DM}$ is the rest mass of DM; $\left<\sigma_{\rm ann} v\right>$ represents the thermally averaged annihilation cross-section; $\Delta^2(z)$ represents the clumping factor; $\mathrm{d}N_{\rm ann}/\mathrm{d}E$ represents the energy spectrum of gamma-rays, indicating the number of photons emitted per unit energy per annihilation for different final states. It can be described as a function of the rest frame energy of the photons and $m_{\rm DM}$. The optical depth $\tau$, accounting for gamma-ray attenuation by the EBL, was taken from \citet{franceschini2008extragalactic}. 

We adopted the energy spectrum of the annihilation process from `A Poor Particle Physicist Cookbook for DM Indirect Detection' (\texttt{PPPC4DMID}, \citealp{cirelli2011pppc}) with electroweak corrections \citep{ciafaloni2011weak}. Our results are presented for three different final state particles: $b\bar{b}$, $\mu^{+}\mu^{-}$ and $\tau^{+}\tau^{-}$. In each of these three scenarios, we assumed that DM annihilates and decays into each final state with a branching ratio of 1. The continuous energy spectrum was obtained through two-dimensional linear interpolation for each state according to the DM mass and gamma-ray energy. 

The clumping factor $\Delta^2(z)$ in Eq. \ref{Eq: w_ann} reflects the concentration of DM within haloes and subhaloes and indicates the density distribution and the aggregation tendencies of DM in the cosmological structure. For the parameters involved in the clumping factor, we used the assumption in \citet{troster2017cross}, which is defined as (\citealp{fornengo2014particle})
\begin{align}
\label{Eq: clumpings}
\Delta^2(z)= \frac{\left<\rho^2_{\rm DM}\right>}{\bar{\rho}^2_{\rm DM}} &= \int_{M_{\rm min}}^{M_{\rm max}} \mathrm{d}M\frac{\mathrm{d}n}{\mathrm{d}M} (M,z)\left[1+b_{\rm sub}(M,z)\right]\notag \\
&\quad\times\int \mathrm{d}^3\textbf {x} \frac{\rho^2(\textbf {x}|M,z)}{\bar{\rho}^2_{\rm DM}},
\end{align}
where $M_{\rm min}$ and $M_{\rm max}$ represent the minimal and maximal halo masses, assumed as $10^{-6}M_{\odot}$ and $10^{18}M_{\odot}$. As in previous works (e.g., \citealp{cuoco2015dark, troster2017cross}), the minimum halo mass $M_{\rm min} = 10^{-6}M_{\odot}$ was chosen to correspond to a typical WIMP free-streaming mass, below which haloes do not contribute effectively to the signal. The uncertainty associated with the selection of $M_{\rm min}$ is explored in appendix B of \citet{paopiamsap2023constraints}. The current mean DM density is denoted by $\rho_{\rm DM}$, and $\mathrm{d}n/\mathrm{d}M$ is the halo mass function \citep{sheth1999large}. The DM density profile within a halo of mass $M$ and at redshift $z$ is denoted as $\rho(\textbf{x}|M,z)$, which is taken from the Navarro-Frenk-White (NFW) profile \citep{navarro1997universal}. The boost factor $b_{\rm sub}$ quantifies the enhancement of the halo emission attributable to the presence of subhaloes. We considered three scenarios with different concentration parameters for subhaloes: \textsc{low} \citep{sanchez2014flattening}, \textsc{mid} \citep{moline2017characterization} and \textsc{high} \citep{gao2012will}.

\begin{figure}
   \centering
   \includegraphics[width=\hsize]{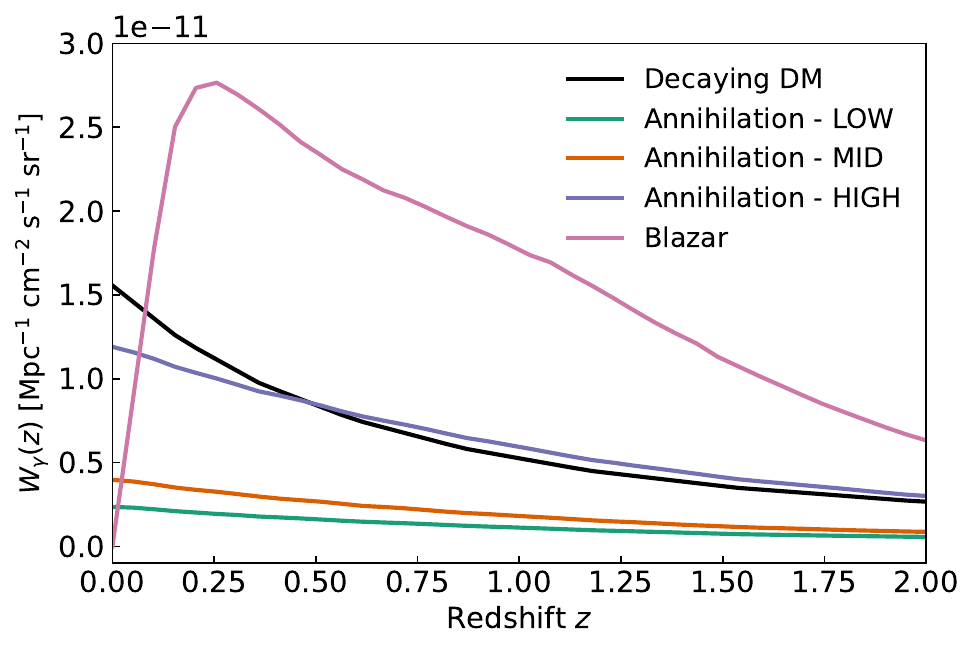}
   \caption{The window functions for gamma-rays produced by DM annihilation, decay, and astrophysical sources across an energy range of $0.5-1000\rm~GeV$. The annihilation process assumes $\left<\sigma_{\rm ann} v\right>=3\times10^{-26} \rm ~cm^{3}s^{-1}$ \citep{steigman2012precise}, where the three scenarios are considered: \textsc{high} (purple), \textsc{mid} (orange), \textsc{low} (green), representing different amounts of DM concentration for subhaloes. The DM decay (black) assumes a particle decay rate of $\Gamma_{\rm dec} = 5\times10^{-28}~\rm s^{-1}$, taken as a representative benchmark consistent with current limits. Both processes are modelled with the assumption of the final state in $b\bar{b}$ pairs and $m_{\rm DM}=100~\rm GeV$ for this figure. The contribution from unresolved blazars, which were taken as the only astrophysical source of the UGRB in this analysis, is shown in pink.}
    \label{Fig: g_window}
    \end{figure}

The process of DM decay involves the spontaneous transition of unstable DM particles into standard model particles, accompanied by the emission of gamma-rays. The window function for this decay process is described as follows (\citealt{ando2006anisotropy, ibarra2013indirect, fornengo2014particle, troster2017cross}):
\begin{align}
\label{Eq: wdec}
W_{\rm g_{\rm dec}}(z, E) = \frac{\Omega_{\rm DM}\rho_c}{4\pi}\frac{\Gamma_{\rm dec}}{m_{\rm DM}} \frac{\mathrm{d}N_{\rm dec}}{\mathrm{d}E}[E(1+z)]\rm e^{-\tau[\it z, E(1+z)]} ,
\end{align}
where $\Gamma_{\rm dec}$ is the decay rate; $\mathrm{d}N_{\rm dec}/\mathrm{d}E$ represents the energy spectrum of the DM decay process. It is analogous to DM annihilation at twice the energy (\citealp{cirelli2011pppc}), which can be expressed as $\mathrm{d}N_{\rm dec}/\mathrm{d}E(E) = \mathrm{d}N_{\rm ann}/\mathrm{d}E(2E)$. The window functions for both the annihilation and the decay of DM are shown in Fig. \ref{Fig: g_window}.

\subsection{Gamma-rays from astrophysical sources}
\label{Sec:astro}

The UGRB is dominated by unresolved astrophysical sources, which constitute the main source of contamination when probing DM signals. The astrophysical gamma-ray background is generally thought to be dominated by contributions from blazars, mAGN, and SFGs. Compared to mAGN and SFGs, blazars are less numerous but produce a higher level of spatial anisotropy at the current level of sensitivity. Consequently, blazars are expected to dominate the angular power spectrum of the UGRB (\citealt{cuoco2012joint, di2014fermi}). The window function from the astrophysical sources is shown as the pink line in Fig.~\ref{Fig: g_window}.

In this analysis, we adopted the assumptions from \citet{korsmeier2022flat}, considering blazars as the only contributors to the unresolved astrophysical gamma-ray background. Specifically, we used an averaged luminosity-dependent density evolution (LDDE) model based on two populations of blazars: BL Lacertae objects and Flat Spectrum Radio Quasars, which are more numerous in the 4GFL catalogue \citep{ajello2015origin}. The window function for astrophysical sources is modelled in the general form:
\begin{align}
\label{Eq: window_astro}
W_{\rm g_{\rm S}}(z, E) = \chi(z)^2\left<f_{\rm S}\right> ,
\end{align}
where $\left<f_{\rm S}\right>$ is the mean flux. For blazars, it can be expressed as:
\begin{align}
\label{Eq: mean_flux}
\left<f_{\rm S}\right>  &= \int^{3.5}_1 \mathrm{d}\Gamma \int^{L_{\rm max}}_{L_{\rm min}} \mathrm{d}L~ \Phi (L, z, \Gamma)\times \frac{\mathrm{d}F}{\mathrm{d}E}(L, z, \Gamma) (1-\Omega (S, \Gamma)) ,
\end{align}
where $\Gamma$ is the photon spectral index, where the integral range follows the assumption in \citet{ajello2015origin}; $L$ is the rest-frame gamma-ray luminosity in the energy range $[0.1, 100]$ GeV; the luminosity bounds are $L_{\rm min} = 10^{43 }\rm erg/s$ and $L_{\rm max} = 10 ^{52}\rm erg/s$ ; $\Phi$ is the gamma-ray luminosity function (GLF), which provides the number density of sources per unit luminosity and per unit co-moving volume at redshift $z$ and photon spectral index $\Gamma$; $\mathrm{d}F/\mathrm{d}E$ is the spectral energy distribution (SED); the term $\Omega$ refers to the Fermi-LAT sensitivity to detect sources.

The GLF term taken from \citet{ajello2015origin} can be decomposed into its expression at $z = 0$ and a redshift-evolution term in the form $\Phi(L,z,\Gamma)=\Phi(L,z=0,\Gamma)\,e(z,L)$, where
\begin{align}
\label{Eq: phi}
\Phi (L, z=0, \Gamma)  &= 
\frac{A}{\ln(10)\, L_\gamma}
\left[
\left( \frac{L_\gamma}{L_*} \right)^{\gamma_1}
+
\left( \frac{L_\gamma}{L_*} \right)^{\gamma_2}
\right]^{-1}  \rm e^{-0.5[\Gamma - \mu(L_\gamma)]^2/\sigma^2}.
\end{align}
%\textcolor{green}{From here [}
The parameters are taken from the LDDE best-fit values presented in Table~1 of \citet{ajello2015origin}.

The SED term is modelled as a power law:
\begin{align}
\label{Eq: sed0}
\frac{\mathrm{d}F}{\mathrm{d}E} = \int dE~ \frac{\mathrm{d}N_{\rm S}}{\mathrm{d}E} \rm e^{-\tau(\it E,z)} 
\end{align}
where the integral range is $[0.1, 100]$ GeV; the term $\rm e^{-\tau(E,z)}$ refers to the attenuation of photons by the EBL; and $\mathrm{d}N_{\rm S}/\mathrm{d}E$ is the observed energy spectrum, which can be described as \citep{ajello2015origin}:
\begin{align}
\label{Eq: sed1}
\frac{\mathrm{d}N_{\rm S}}{\mathrm{d}E} = K\left[\left(\frac{E}{E_\mathrm{b}}\right)^{1.7}+\left(\frac{E}{E_\mathrm{b}}\right)^{2.6}\right]^{-1}.
\end{align}
We note that the GLF term was defined with luminosities in the rest frame energy range of $[0.1, 100]$ GeV. When predicting the unresolved gamma-ray flux, we extrapolated the corresponding source spectra to our analysis range of $[0.5, 1000]$ GeV, including attenuation by the EBL, in order to match the gamma-ray energy bins used in the cross-correlation. Similar methods were applied to the flux regime, where GLF models calibrated on resolved Fermi-LAT sources were extrapolated to the unresolved population (e.g. \citealt{manconi2020testing, korsmeier2022flat}).

From simulations, the relation between break energy $E_b$ and the LAT-measured power-law photon index can be approximated as $\text{log} E_b (\text{GeV})\approx 9.25 -4.11 \Gamma $. The normalisation factor $K$ is given by \citep{zhang2022constraints}
\begin{align}
\label{Eq: K}
K = L \left[4\pi d_L^2 k\int \mathrm{d}E~ E \frac{\mathrm{d}N_{\rm S}}{\mathrm{d}E} (K=1)\right]^{-1}, 
\end{align}
where $k$ is the $K$-correction term, obtained as \citep{wang2015gamma}:
\begin{align}
\label{Eq: kc}
k = \left[\int^{E_{\rm max}/(1+z)}_{E_{\rm min}/(1+z)}\mathrm{d}E~ E \frac{\mathrm{d}N_{\rm S}}{\mathrm{d}E}\right] \left[\int^{E_{\rm max}}_{E_{\rm min}}\mathrm{d}E~ E \frac{\mathrm{d}N_{\rm S}}{\mathrm{d}E}\right]^{-1}.
\end{align}
The term $\Omega$ in Eq. \ref{Eq: mean_flux} refers to the Fermi-LAT sensitivity and is modelled as a step function \citep{manconi2020testing}
\begin{align}
\label{Eq: omega}
\Omega(\Gamma) = \Theta[S_{E, 100} - S_{\rm thr}(\Gamma)],
\end{align}
where sources with a given spectral index $\Gamma$ are regarded as undetected ($\Omega = 0$) if their flux falls below the detection threshold $S_{\rm thr}(\Gamma)$ of the fourth Fermi-LAT source catalogue (4FGL, \citet{abdollahi2020fermi}); and $S_{E, 100} = \int^{100 \rm ~GeV}_{0.1 ~\rm GeV} \mathrm{d}E ~E \frac{\rm d\it F}{\rm d\it E}$. 

\begin{figure}
   \centering
   \includegraphics[width=\hsize]{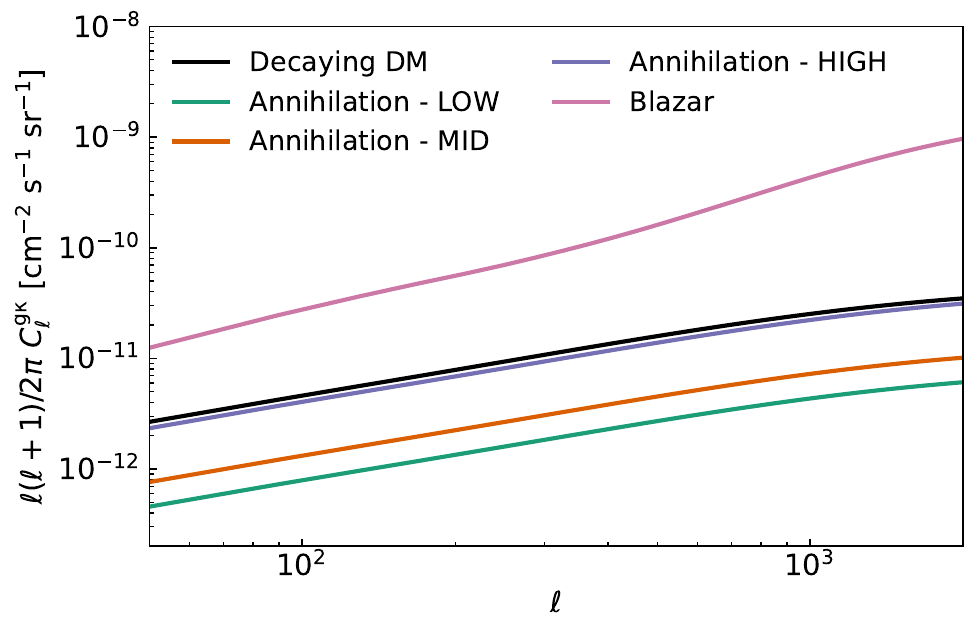}
   \caption{Model of $C^{\rm g\kappa}_{\ell}$ for three clumping cases of DM annihilation, decaying DM, and astrophysical background. The models and formatting are the same as in Fig. \ref{Fig: g_window}. The models assumed the redshift $n(z)$ for $z$ in the range of $[0.1, 2.0]$ and energy range from 0.5 to 1000 GeV. The effect of the Fermi-LAT PSF was not considered. The blazar component shown in pink represents the unresolved blazar contribution to the astrophysical background adopted in this analysis.}
    \label{Fig: cells}
    \end{figure}

\subsection{Halo model}
\label{Sec:halo}

The 3D power spectrum can be effectively described using the DM halo model \citep[see][and references therein]{asgari2023OJAp....6E..39A}. Within this formalism, the power spectrum $P_{uv}(k)$ is decomposed into two components: the two-halo term, representing the correlation between different haloes; and the one-halo term, representing the correlations within the same halo:
\begin{gather}
\label{Eq: 1+2halo}
P_{uv}(k) = P_{uv}^{2\mathrm{h}}(k) + P^{1 \mathrm{h}}_{uv}(k) .
\end{gather}

The general 3D cross-power spectrum of one- and two-halo terms, related to the fields $u$ and $v$ is given by:
\begin{gather}
\label{Eq: 3d_spec_halo}
P^{1h}_{uv} (k) = \int_{M_{\rm min}}^{M_{\rm max}} \mathrm{d}M \frac{\mathrm{d}n}{\mathrm{d}M}(M)\left<\hat u(k|M)\hat v(k|M)\right>,\\
P^{2h}_{uv} (k) = \left<b\hat u\right>(k)\left<b\hat v\right>(k) P^{\rm lin} (k),\\
\left<b\hat u\right>(k) = \int_{M_{\rm min}}^{M_{\rm max}} \mathrm{d}M \frac{\mathrm{d}n}{\mathrm{d}M}b_{\rm h}(M)\left<\hat u(k|M)\right>, 
\end{gather}
where $P^{\rm lin}$ is the linear power spectrum; $b_\mathrm{h}$ is the linear bias; $\hat u(k|M)$ and $\hat v(k|M)$ represent the Fourier transforms of the halo profile for $u$ and $v$ with mass $M$.

\begin{figure}
   \centering
   \includegraphics[width=\hsize]{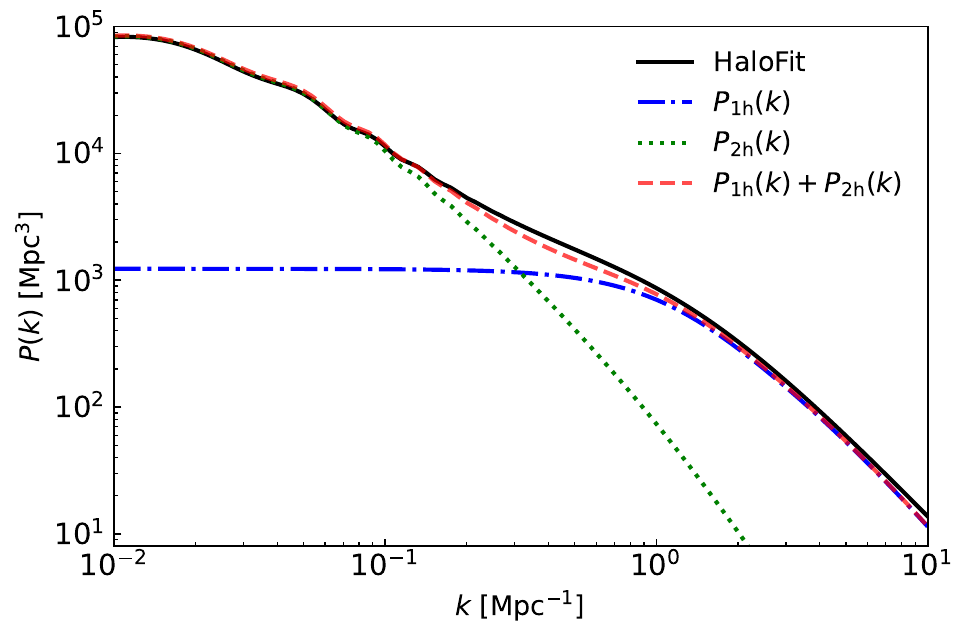}
   \caption{Halo model of the power spectrum of gamma-ray contributions from DM annihilation and shear cross-correlation at $z=0$. The blue dash–dotted and green dotted curves show the one-halo and two-halo terms, respectively; the red dashed curve is their summation, and the black solid curve shows the power spectrum fitted with \texttt{halofit} at $z=0$. }
    \label{Fig: halo}
    \end{figure}

We calculated the cross angular power spectrum of convergence and gamma-rays emitted from DM with the Core Cosmology Library (CCL, \citealp{chisari2019core}). The matter power spectrum was estimated via the \texttt{CAMB} library \citep{lewis2000efficient}, where the non-linear part of the spectrum was fitted using \texttt{halofit} (\citealp{smith2003stable, takahashi2012revising}). The prediction of the angular power spectrum without the Fermi-LAT PSF correction (as discussed in Sect. \ref{Sec:fermilat}) is demonstrated in Fig. \ref{Fig: cells}.
We note that \texttt{halofit} predicts a slightly higher power spectrum than a halo model prediction in the 1– and 2-halo transition regime, as shown in Fig.~\ref{Fig: halo}, which would propagate into mildly tighter DM upper limits.

For the analysis of cross-correlating the cosmic shear and unresolved astrophysical sources, we adopted the one- and two-halo models with the window function detailed in Sect.~\ref{Sec:astro}. The 3D cross-power spectrum of shear and astrophysical sources can be expressed as:
\begin{align}
\label{Eq: 3d_spec_astro}
P^{1\rm h}_{\rm g{\rm S}\kappa} (k, z) &= \int_{L_{\rm min}}^{L_{\rm max}} \mathrm{d}L \frac{\Phi(L, z)}{\left<f_{\rm S}\right>} \frac{\mathrm{d}F}{\mathrm{d}E}(L, z)\hat u_{\kappa}(k|M(L, z), z),\\
P^{2\rm h}_{\rm g{\rm S}\kappa} (k, z) &= \left[\int_{L_{\rm min}}^{L_{\rm max}}\mathrm{d}L ~b_{\rm S} (L, z) \frac{\Phi(L, z)}{\left<f_{\rm S}\right>}\frac{\mathrm{d}F}{\mathrm{d}E}(L, z) \right] \notag \\
&\quad \times \left[\int_{M_{\rm min}}^{M_{\rm max}}\mathrm{d}M \frac{\mathrm{d}n_\mathrm{h}}{\mathrm{d}M}b_\mathrm{h}(M,z)\hat{u}_{\kappa}(k | M,z) \right] \notag \\
&\quad\times P^{\rm lin}(k, z),
\end{align}
where $b_{\rm S}$ is the bias of gamma-ray astrophysical sources relative to the matter distribution, which can be computed in terms of the halo bias $b_{\rm S} (L, z) = b_\mathrm{h}(M(L, z))$ as assumed in \citet{2015JCAP...06..029C}, and the linear bias $b_\mathrm{h}$ is taken from the model of \citet{tinker2010large} ; $\Phi$ is the GLF of blazars; $\mathrm{d}F/\mathrm{d}E$ is the differential photon flux; and $\left<f_{\rm S}\right>$ is the mean flux of blazars.
 
We modelled the angular cross-power spectrum of astrophysical sources from the UGRB and weak lensing signals also using CCL. 
For the transformation of the gamma-ray luminosity of a blazar to the mass of its host DM halo, we adopted the `Model B', which corresponds to an intermediate scaling between halo mass and blazar luminosity, following the relation introduced by \citet{camera2013novel}:
\begin{align}
\label{Eq: L-dm}
M = 10^{13} M_{\odot}\left(\frac{L}{10^{44}\rm ~erg~s^{-1}}\right)^{1.7}.
\end{align}

\section{Data}
\label{Sec:Data}

\subsection{KiDS}
\label{Sec:kids data}

We used the fifth data release from the Kilo-Degree Survey (KiDS-Legacy, \citealp{Wright2024A&A...686A.170W}) for the weak gravitational lensing data sample. KiDS is a wide-field optical imaging survey conducted using the OmegaCAM camera on the VLT Survey Telescope (VST, \citealt{capaccioli2011vlt}), specifically designed for weak lensing applications. 
The footprint of the KiDS can be divided into two patches: KiDS-N, covering regions along the equator, and KiDS-S, covering the southern sky. In conjunction with the infrared data from the VISTA Kilo-degree INfrared Galaxy survey (VIKING, \citealp{edge2013vista}), the KiDS-Legacy footprint spans 1347 $\rm deg^2$, with an effective area of 967.4 $\rm deg^2$ after masking. The KiDS-Legacy dataset provides photometry in nine optical and near-infrared bands: $ugriZYJHK_{\rm s}$, along with a second $i$‑band pass and extra 23 $\rm deg^2$ of KiDS-like calibration observations of deep spectroscopic surveys  \citep{wright2025arXiv250319441W}, allowing for reliable photometric redshift estimation and redshift distribution calibration in KiDS-Legacy.

The galaxy shape measurements in KiDS-Legacy were performed with \textit{lens}fit (\citealp{miller2007MNRAS.382..315M, miller2013bayesian}), and were calibrated with SKiLLS simulations as detailed in \citet{Li2023A&A...670A.100L}, which demonstrated that residual shear-related systematic errors are negligible after calibration. 
The robustness of the KiDS shear measurements were further validated by an
independent \textsc{MetaCalibration}-based analysis \citep{Yoon2025arXiv251001122Y}.
Our study selected galaxies from the gold-selected sample of KiDS-Legacy, which offers a high-accuracy redshift distribution. 
We conducted a tomographic cross-correlation analysis by dividing the sample into six photometric redshift bins: $(0.10, 0.42], ~(0.42, 0.58], ~(0.58, 0.71], ~(0.71, 0.90], ~(0.90, 1.14]$ and $(1.14, 2.00]$, based on the Bayesian photometric redshift $z_B$ inferred from the \textsc{bpz} algorithm \citep{benitez2000bayesian}.
The redshift distributions $n(z)$ were estimated and calibrated via a combination of self-organising map (SOM) colour-based direct calibration, and cross-correlation methods using nine-band KiDS+VIKING photometry, and deep spectroscopic reference samples as detailed in \citet{2025arXiv250319440W}. We summarise the shape noise and effective galaxy number density of KiDS-Legacy for each tomographic bin in Table \ref{Tab:shape_noise}.

\begin{table}[]
\centering
\caption{Information on the KiDS-Legacy gold-selected sample in the six tomographic redshift bins \citep{wright2025arXiv250319441W}. The effective number density of the sample is denoted by $n_{\rm eff}$, and $\sigma_e$ represents the measured ellipticity dispersion per component.}
\renewcommand{\arraystretch}{1.1}
\begin{tabular}{ccccc}
\hline
\hline
Bin & $z_B$ range    & Mean $z$ & $n_{\rm eff}~[\rm arcmin^{-2}]$ & $\sigma_e$ \\ \hline
1   & {(}0.10, 0.42{]} & 0.335     & 1.77                            & 0.28      \\
2   & {(}0.42, 0.58{]} & 0.477     & 1.65                            & 0.27      \\
3   & {(}0.58, 0.71{]} & 0.587     & 1.50                            & 0.29      \\
4   & {(}0.71, 0.90{]} & 0.789     & 1.46                            & 0.26      \\
5   & {(}0.90, 1.14{]} & 0.940     & 1.35                            & 0.28      \\ 
6   & {(}1.14, 2.00{]} & 1.224     & 1.07                            & 0.30      \\ 
1-6 & {(}0.10, 2.00{]} & 0.680     & 8.79                            & 0.28      \\ \hline
\end{tabular}%
\vspace{4pt}
\label{Tab:shape_noise}
\end{table}

\subsection{Fermi-LAT}
\label{Sec:fermilat}

We analysed 15 years of observational data from the Fermi-LAT, spanning up to 3 July 2023. The Fermi-LAT has been surveying the gamma-ray sky since 2008 \citep{atwood2009large}. Its mission is to conduct long-term, high-sensitivity observations of celestial sources, covering an energy range from $\sim 20~ \rm MeV$ to $>300~\rm GeV$. The LAT is a wide field-of-view imaging gamma-ray telescope with a large effective area, combined with excellent energy and angular resolution.  

In this study, we used the Pass 8 event data and processed the data preparation with \textsc{Fermi Science Tools} version 2.2.0\footnote{\url{https://github.com/fermi-lat/Fermitools-conda}}. Our selection criteria included \texttt{ultracleanveto} photons and applied cuts on the data for higher quality with filters \texttt{DATA\_QUAL>0 \&\& LAT\_CONFIG==1}. Photons with the lowest PSF quartile (PSF0) were excluded, and the surviving events were binned into 100 logarithmically spaced bins between 0.5 and 1000 GeV. The generated photon counts and exposure maps were projected in \textsc{HEALPix}\footnote{\url{https://healpix.sourceforge.net}} \citep{gorski2005healpix} format with \texttt{nside=1024}, corresponding to a pixel size of approximately $3.4 ~\rm arcmin$. Intensity maps were obtained by dividing the counts by the exposure, and were subsequently co-added into the analysis bins as described below. The process is facilitated by the \texttt{healpy} package \citep{Zonca2019}.

\begin{figure}
   \centering
   \includegraphics[width=\hsize]{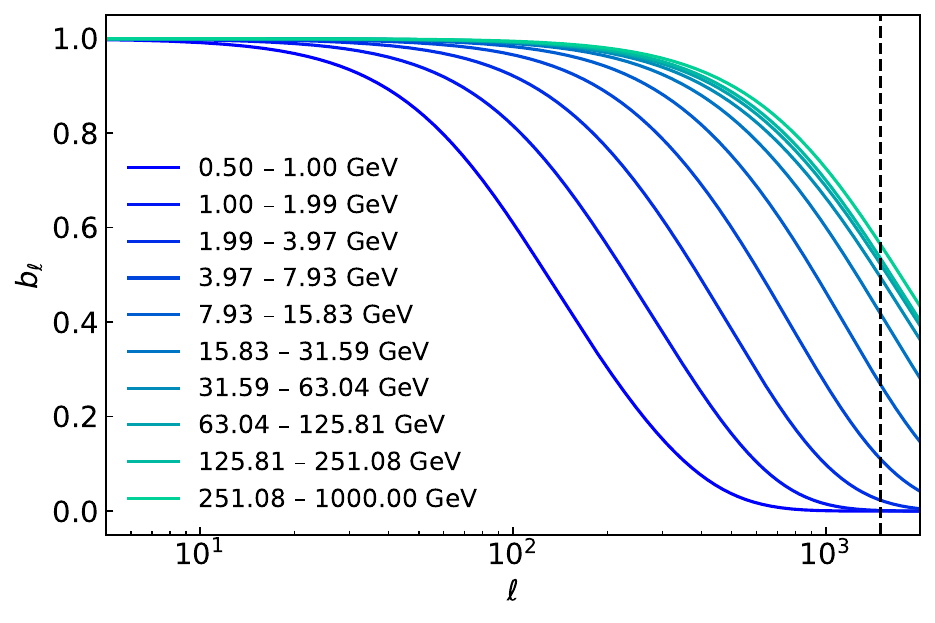}
   \caption{The Fermi PSF in harmonic space $b_\ell$ in ten different energy bins. It shows an inverse relationship between the energy and the suppression of the PSF effect. The black dashed line represents $\ell = 1500$, referring to the upper limit of $\ell$ in the cross-power spectrum measurement.}
    \label{Fig: psf}
    \end{figure}

To isolate the UGRB, we applied masks for both diffuse Galactic emission and point sources. We masked sky regions where the fiducial Galactic diffuse emission model (\texttt{gll\_iem\_v07}\footnote{\url{https://fermi.gsfc.nasa.gov/ssc/data/analysis/software/aux/4fgl/gll_iem_v07.fits}}) exceeds three times the intensity of the isotropic template. All gamma-ray sources from the 4FGL-DR3 catalogue\footnote{\url{https://fermi.gsfc.nasa.gov/ssc/data/access/lat/12yr_catalog/gll_psc_v31.fit}} were then masked. The mask size was set according to the LAT point spread function (PSF), which describes the energy-dependent smearing of photon directions due to the finite angular resolution of the instrument.
We analysed the PSF in each energy bin using the \texttt{P8R2\_SOURCE\_V6} Instrument Response Functions (IRFs). 
In harmonic space, the real-space PSF is represented by the beam window $b_{\ell}$, which attenuates the observed angular power spectrum, particularly at high multipoles (see Fig.~\ref{Fig: psf}). 
For each energy bin, we set the source mask radius to be twice the LAT 68\% containment radius, derived from the corresponding PSF parameterisation of the response functions. To assess the effectiveness of the point source masking we adopted, we estimated the fraction of point source flux leakage. Even in the lowest-energy bin ($0.5$–$1$ GeV), which suffered most from the PSF; the leakage fraction is $\lesssim 3\%$, indicating that residual contamination from resolved sources was negligible for our analysis.

After applying the masks, we refitted the normalisations of the Galactic diffuse and isotropic templates in each of the 100 fine energy bins by maximising a Poisson likelihood. The residual photon counts were then corrected for the LAT exposure and combined into ten energy bins covering $[0.5, 1000]$ GeV, with edges [0.5, 1.0, 1.99, 3.97, 7.93, 15.83, 31.59, 63.04, 125.81, 251.08, 1000]~$\mathrm{GeV}$,  which were used in the subsequent cross-correlation analysis.

\section{Methods}
\label{Sec:Methods}

In this section, we describe the methodologies we employed to calculate the cross-correlation between the galaxy lensing data and the gamma-ray intensity map, to estimate the covariance, and to constrain parameters related to DM.

\subsection{Power spectrum estimation}
\label{Sec:PS estimate}

In this study, we created maps and masks for both KiDS-Legacy and Fermi-LAT data to estimate the cross-correlation angular power spectrum of weak lensing and gamma-ray  intensity. We used the \textsc{MASTER} algorithm \citep{hivon2002master} implemented in \texttt{NaMaster}\footnote{\url{https://github.com/LSSTDESC/NaMaster}} \citep{alonso2019unified} to compute the cross power spectra. \texttt{NaMaster} provides a general framework to estimate pseudo-$C_\ell$ on masked fields with arbitrary spin and contaminants, and to deconvolve mask-induced mode coupling via the coupling matrix.

For the weak lensing component of the KiDS-Legacy data, we created \textsc{HEALPix} maps for different redshift bins with \texttt{nside=1024} covering the footprint of the KiDS-Legacy gold-selected sample catalogue. We took the weighted average of the galaxy ellipticities for each pixel, where the weights were determined by the shape measurement uncertainties, forming the spin-2 field \{$\gamma_1, \gamma_2$\}. The corresponding masks, matching the coverage and resolution of the maps, were generated as a sum of shape measurement weights for each pixel.

For the Fermi-LAT gamma-ray data, we constructed the gamma-ray intensity maps for different energy bins as described in Sect.~\ref{Sec:fermilat}. The mask used for the gamma-ray data was binary (1 for the unmasked regime, and 0 for the masked regime) and energy-dependent, with point sources and Galactic diffuse emission removed.

Using the three fields, \{$\rm g, \gamma_1, \gamma_2$\}, we applied \texttt{NaMaster} to estimate the cross-power spectra between the weak lensing E/B-modes and the gamma-ray intensity $I_\gamma$. 
The E-mode is the lensing-induced component and thus carries the cosmological shear signal, which can correlate with gamma-ray emission from astrophysical sources, as well as any potential DM contribution.
The B-mode of the shear is obtained by rotating galaxy shapes by $45^\circ$.
Since gravitational lensing is parity-conserving, pure gravitational lensing is
expected to produce only E-mode patterns, with no B-modes in the absence of
systematic effects or intrinsic alignments. The B-mode therefore provides a
valuable null test and diagnostic of residual systematics.
The E- and B-modes are analogous to the decomposition used in CMB polarisation studies (see, e.g., \citealp{PhysRevD.55.7368}), where E/B-modes represent the curl-free and divergence-free components, respectively.

The resultant cross-power spectrum was organised into five linearly spaced multipole bins spanning from 200 to 1500. 
The upper limit was set by the Fermi–LAT PSF, which suppresses power on small angular scales (high $\ell$; see Fig.~\ref{Fig: psf}), whereas the lower limit was chosen considering the survey sky coverage and residuals from foreground subtraction on large angular scales.

\subsection{Covariance}
\label{Sec:cov}

We combined two approaches to estimate the covariance of the angular cross-power spectra between tomographic weak-lensing data and energy-binned gamma-ray intensity maps: a weighted jackknife and a Gaussian covariance computed with \texttt{NaMaster}.

Jackknife resampling is a non-parametric technique that estimates covariance by systematically excluding one or more tiles of observations from the dataset and recalculating the estimate across all subsets. 
In our analysis, we employed the delete-one weighted jackknife resampling to estimate the covariance of the cross-correlation between weak lensing data and gamma-rays. 

For each redshift–energy bin, we divided the corresponding footprint into 50 approximately equal-area sub-regions using the \textsc{HEALPix} scheme.
The $k$-th sub-region was removed from the sample, and we recalculated the pseudo-$C_{\ell}$ of the remaining parts for all realisations. Realisations were weighted by the sum of galaxy shape measurement weights in the retained area. The covariance of the measurement was computed as follows \citep{mohammad2022creating}:
\begin{align}
\label{Eq: cov_jn}
{\bf C}^{\rm JK}(C_\ell^{\rm g\kappa}, C_{\ell'}^{\rm g\kappa})
=
(n_s - 1)
\frac{
\sum_{k}^{n_s}
w_k
\left(C_{\ell,k}^{\rm g\kappa}-\left< C_{\ell}^{\rm g\kappa}\right>\right)
\left(C_{\ell',k}^{\rm g\kappa}-\left< C_{\ell'}^{\rm g\kappa}\right>\right)
}{
\sum_k w_k}.
\end{align}
where $n_s = 50$ is the number of subsamples used in the jackknife resampling, $w_k$ is the weight for the $k$-th patch, $C_{\ell, k}^{\rm g\kappa}$ is the cross-power spectrum between convergence and gamma-rays for the $k$-th patch, and $\left<C_{\ell, k}^{\rm g\kappa}\right>$ is the weighted average of the cross-angular power spectrum for all jackknife subsamples.
Because the gamma-ray masks were energy dependent, the jackknife procedure was applied independently to each redshift–energy bin, and the resulting estimates were used only for the corresponding diagonal covariance blocks. Cross-bin correlations were supplied by the Gaussian covariance described below.

Gaussian covariance estimation with \texttt{NaMaster} was employed under the Gaussian-field assumption and accounted for the mode-coupling induced by the survey masks through coupling matrices. 
In practice, for each pair of fields $\{\rm g,\kappa\}$, we first measured the coupled pseudo-$C_\ell$ $\tilde C_\ell^{\rm g\kappa}$ from the masked maps, and normalised them by the effective overlap of the corresponding masks:
\begin{align}
C_{\ell}^{g\kappa, \rm Cov} \simeq\frac{\hat{C}^{\rm g\kappa}_{\ell}} {\langle w_{\rm g} w_\kappa \rangle},
\end{align}
where $C_{\ell}^{\rm g\kappa, \rm Cov}$ is the power spectrum between the gamma-ray field $g$ and the convergence field $\kappa$ used to estimate the covariance matrix. The term $\hat{C}^{\rm g\kappa}_{\ell}$ is the corresponding pseudo-$C_\ell$. The gamma-ray masks $w_\gamma$ are binary maps, and the convergence masks $w_\kappa$ correspond to the weighting maps of the galaxy shape measurements. The factor $\langle w_{\rm g} w_\kappa \rangle $ represents the average product of the masks over all pixels.

We assessed the consistency of the \texttt{NaMaster} Gaussian covariance by comparing its diagonal terms with jackknife estimates across all tomographic and energy bins, and found a good agreement. Therefore, to construct the fiducial covariance used in the likelihood, we combined both estimators: we used the jackknife variance and combined it with the correlation matrix that captures mode coupling and cross-bin correlations from \texttt{NaMaster}. Specifically, we rescaled the \texttt{NaMaster} covariance to match the jackknife variances on the diagonal \citep{koukoufilippas2020MNRAS.491.5464K},
\begin{align}
\mathbf{C}^{\rm Comb}_{ij}
= \mathbf{C}^{\rm NMT}_{ij}\sqrt{\frac{\mathbf{C}^{\rm JK}_{ii}\mathbf{C}^{\rm JK}_{jj}}
{\mathbf{C}^{\rm NMT}_{ii}\mathbf{C}^{\rm NMT}_{jj}}}\,
\end{align}
where $\mathbf{C}^{\rm JK}$ and $\mathbf{C}^{\rm NMT}$ denote the jackknife and \texttt{NaMaster} Gaussian covariance matrices, respectively. 
The correlation matrix corresponding to the combined covariance is shown in Appendix~\ref{App:CovMatrix}.

For the forecast covariance, we assumed Gaussian-distributed fields and accounted for mask-induced mode-coupling using \texttt{NaMaster}. In this case, we modelled power spectra $\hat{C}^{\kappa\kappa'}_\ell$, $\hat{C}^{\rm gg'}_\ell$ and $\hat{C}^{\rm g\kappa}_\ell$ as inputs, where $\hat{C}^{\kappa\kappa'}_\ell$ and $\hat{C}^{\rm gg'}_\ell$ denote the estimated auto power spectra of convergence $\kappa$ and gamma-ray $\rm g$, respectively. $\hat{C}^{\rm g\kappa}_\ell$ denotes the cross power spectrum as defined in Sect.~\ref{Sec:Models}.

The lensing auto power spectrum $\hat{C}^{\kappa\kappa'}_\ell$ was modelled as the combination of the cosmic shear signal and shape noise: 
\begin{align}
\label{Eq: cov_kk}
\hat{C}^{\kappa\kappa'}_\ell = {C}^{\kappa\kappa'}_\ell  + \delta_{\kappa\kappa'}\frac{\sigma_e^2}{n_{\rm eff}},
\end{align}
where ${C}^{\kappa\kappa'}_\ell$ is the cosmic shear signal computed for each tomographic bin; $\sigma_e^2$ and $n_{\rm eff}$ are the dispersion of the ellipticities, and effective galaxy number density of the survey, which only contribute to the auto spectra terms.

\begin{figure*}
  \centering
   \includegraphics[width=\hsize]{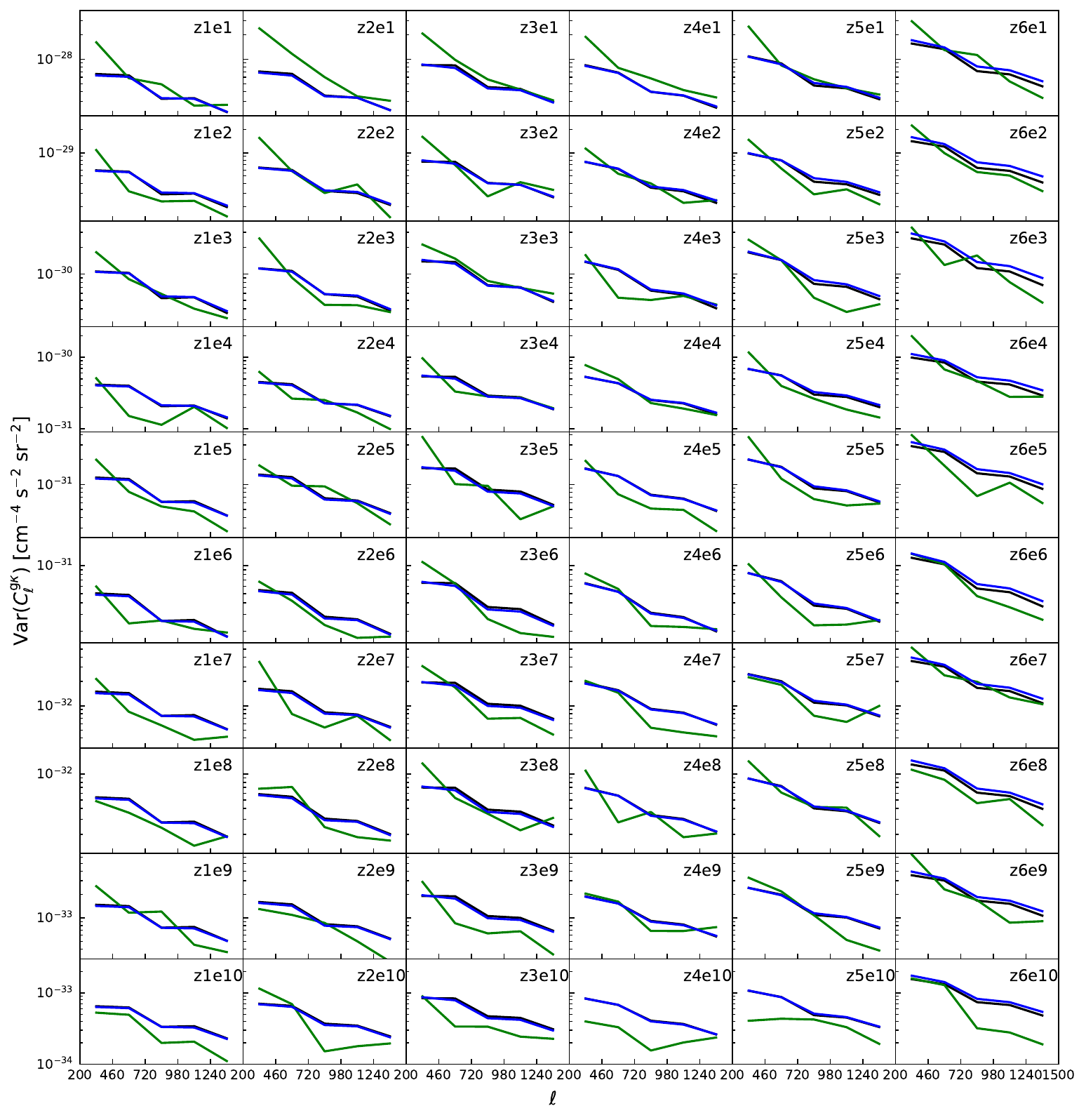}
     \caption{Comparison of diagonal variances for the gamma–shear cross-power spectra. Each subplot corresponds to one tomographic redshift bin (columns, from z1 to z6) and one Fermi–LAT energy bin (rows, from e1 to e10). The diagonal elements $\mathrm{Var}(C_\ell^{g\kappa})$ are shown in the five multipole bins used in the measurement, with weighted jackknife estimates in green, \texttt{NaMaster}-analytic Gaussian covariance in blue, and \texttt{NaMaster}-measured covariance in black.}
         \label{Fig:var}
   \end{figure*}

The gamma-ray power spectrum $\hat{C}^{\rm gg'}_\ell$ was estimated from the gamma-ray flux maps used in the cross-correlation. We measured the auto- and cross-angular spectra among the ten energy bins using \texttt{NaMaster} in 15 logarithmically spaced $\ell$-bins between 30 and 2000. We fitted the measured spectra with the form
\begin{align}
\label{Eq: cov_gg}
\hat{C}^{\rm gg}_\ell = {C}_{\rm Poisson}  + c~\ell^\alpha,
\end{align}
where the Poisson term ${C}_{\rm Poisson}$ captures the photon shot noise, and $c$ and $\alpha$ account for a power-law contribution for large scales. For the auto-spectra, we found that the Poisson term dominates and the best-fitting amplitude $c$ was consistent with zero, while for cross-spectra the Poisson term vanished and only the power-law component was retained. We compare the diagonal terms (variances) of the jackknife (green), the \texttt{NaMaster}-analytic (blue) and \texttt{NaMaster}-measured (black) estimates in Fig.~\ref{Fig:var}.

\subsection{Likelihood}
\label{Sec:likelihood}

We constrained the model parameters $\boldsymbol{p} = \{\left<\sigma_{\rm ann}v\right>, \Gamma_{\rm dec}\}$ for each DM mass bin as described in Sect.~\ref{Sec:Models}, assuming the measured power spectra follow a Gaussian likelihood:
\begin{equation}
\chi^{2} = -2 \ln L(\boldsymbol{d}\,|\,\boldsymbol{p})
= \bigl[ \boldsymbol{d} - \boldsymbol{\mu}(\boldsymbol{p}) \bigr]^{\mathrm{T}}
\mathbf{C}^{-1}
\bigl[ \boldsymbol{d} - \boldsymbol{\mu}(\boldsymbol{p}) \bigr],
\label{Eq: chi2}
\end{equation}
where $\boldsymbol{d}$ is the data vector, comprising $6\times10$ cross-correlations with six redshift bins in weak lensing data and ten energy bins in gamma-ray data; $\boldsymbol{\mu}(\boldsymbol{p})$ is the theoretical cross-correlation predicted by the model parameters; $\rm \bf C^{\rm -1}$ is the inverse of the covariance matrix. Uniform priors are assigned for the model parameters $\boldsymbol{p}$. 

Given that our data lack the constraining power to fully determine the DM parameters, we derived the 95\% confidence upper limit constraints on $\left<\sigma_{\rm ann}v\right>$ and $\Gamma_{\rm dec}$ from the contours of the likelihood surface. We followed the assumption from \citet{troster2017cross}, and for a specified confidence interval $p_{\rm c}$, the contours were defined by the parameter sets $\boldsymbol{d}$, where
\begin{align}
\label{Eq: chi2_p}
\chi^{2}(\boldsymbol{d}, \boldsymbol{p}) =
\chi^{2}(\boldsymbol{d}, \boldsymbol{p}_{\rm ML})
+ \Delta \chi^{2}(p_{\rm c}),
\end{align}
where $\boldsymbol{p}_{\rm ML}$ is the maximum likelihood estimation of the parameters, $\chi^2$ is calculated from Eq. \ref{Eq: chi2}, and the $\Delta{\chi^2} (p_{\rm c})$ corresponds to the quantile function of the $\chi^2$ distribution. For $2\sigma$ contours, $\Delta_{\chi^2}(0.95) = 6.18$. We calculated the limits on $\left<\sigma_{\rm ann}v\right>$ and $\Gamma_{\rm dec}$ for different final states of standard model particles with $m_{\rm DM}$ in logarithmically spaced values in the range $\left[10, 1000\right]$ GeV.

\section{Results}
\label{Sec:Results}

\subsection{Cross-correlation measurement}
\label{subsec: results_cross-corr}

\begin{figure*}
  \centering
   \includegraphics[width=\hsize]{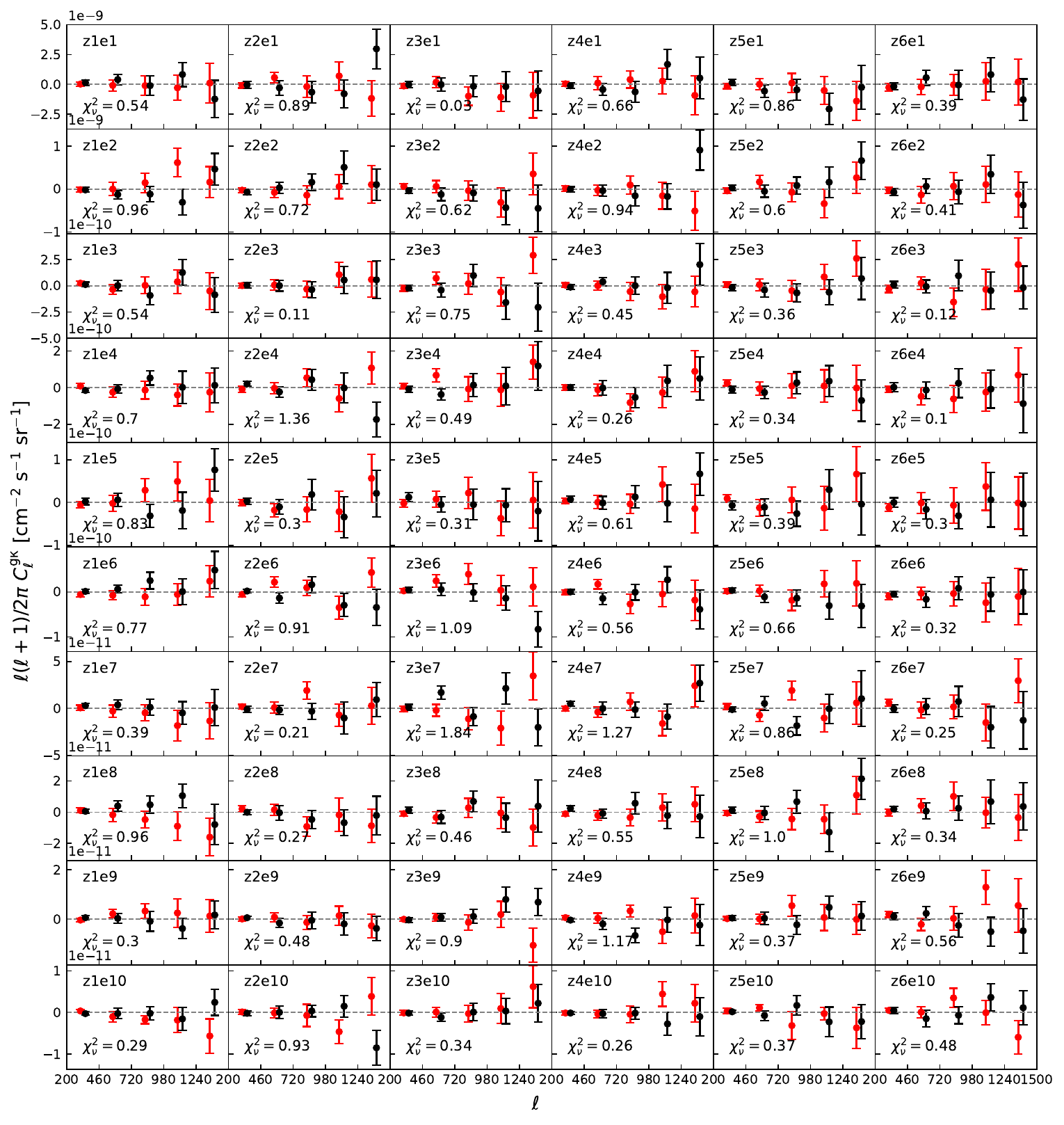}
     \caption{Cross angular power spectrum $C_\ell^{\rm g\kappa}$ between the Fermi-LAT gamma-ray intensity map and KiDS-Legacy weak lensing data, using a hybrid covariance (jackknife diagonals plus off-diagonals described in Sect.~\ref{Sec:cov}). E- and B-mode cross-spectra are shown as black and red points, respectively. Panels are labelled by the tomographic redshift bin $z_i$ and gamma-ray energy bin $E_j$. The reduced $\chi^2_\nu$ ($\nu= 5$) with respect to null signal for the E-mode in each bin is also provided in each subplot.}
         \label{Fig:pcl}
   \end{figure*}

We present our measurements of the cross-correlation angular power spectrum between the Fermi-LAT gamma-ray intensity maps (across ten energy bins) and KiDS-Legacy weak lensing data (across six redshift bins), utilising a combined covariance, as detailed in Sect. \ref{Sec:cov}. We show the results in Fig. \ref{Fig:pcl}, with black and red points representing the cross-correlation of gamma-rays with the components of the E- and B-modes of shear, respectively.

The $\chi^2_\nu$ values with respect to the null signal with degrees of freedom $\nu = 5$ are illustrated in subplots for all bins. 
The global $\chi^2$ for the E-mode signal is 176.51, evaluated over 300 data points (corresponding to a p-value close to unity), which indicates that there is no significant detection of cross-correlation signals.
We note that this $\chi^2$ is below the degrees of freedom, suggesting that the covariance could be somewhat overestimated.
As discussed in Sect.~\ref{Sec:cov}, the jackknife, \texttt{NaMaster}–measured, and \texttt{NaMaster}–analytic covariance estimates yield consistent diagonal variances, indicating internal consistency among the different estimation methods.
However, all three approaches assume isotropic and spatially uniform noise, which is not strictly valid for the Fermi–LAT intensity maps.
Because the gamma-ray maps are dominated by photon shot noise that depends on the local exposure time, and since the exposure varies across the sky, the effective noise level is spatially inhomogeneous.
This can lead to mildly overestimated variances, and hence a lower global $\chi^2$.
We tested this by splitting the KiDS footprint into North and South patches. 
We removed the area of KiDS–N with the strongest exposure variations to make both regions have comparable effective sky coverage. As a side effect, this also makes KiDS–N more uniform in exposure than KiDS–S, which helps isolate the effect of inhomogeneous photon noise.
The result shows that KiDS-S, which exhibits stronger Fermi exposure variations, yields a lower $\chi^2$ (163.82 in KiDS–S and 211.30 in KiDS–N, among 300 data points). This is consistent with the interpretation that anisotropic photon noise leads to a conservative covariance estimate.
Overall, these results are consistent with the non-detection of signals across all cross angular power spectrum measurements for the six redshift bins and ten energy bins. Furthermore, for B-modes, the $\chi^2$ value is 185.61 across the same ten energy and six redshift bins.

\subsection{Constraints on DM parameters}
\label{subsec: results_constraints_dm} 

We applied the formalism detailed in Sect.~\ref{Sec:likelihood} to derive constraints on the DM thermally averaged annihilation cross-section $\left<\sigma_{\rm ann} v\right>$ and decay rate $\Gamma_{\rm dec}$ for three final states: $b\bar{b},~ \mu^+\mu^-$, and $\tau^+\tau^-$, each state with a branching ratio of 1. 
In the modelling, we accounted for the blazar contribution to the UGRB as described in Sect.~\ref{Sec:astro}.
Although the background of astrophysical sources dominates the mixed UGRB, the improvements in constraints after excluding blazar effects remain limited. Due to the non-detection of cross-correlation signals, we can only provide 95\% upper bounds, without establishing both upper and lower limits.

\begin{figure*}
  \centering
   \includegraphics[width=\hsize]{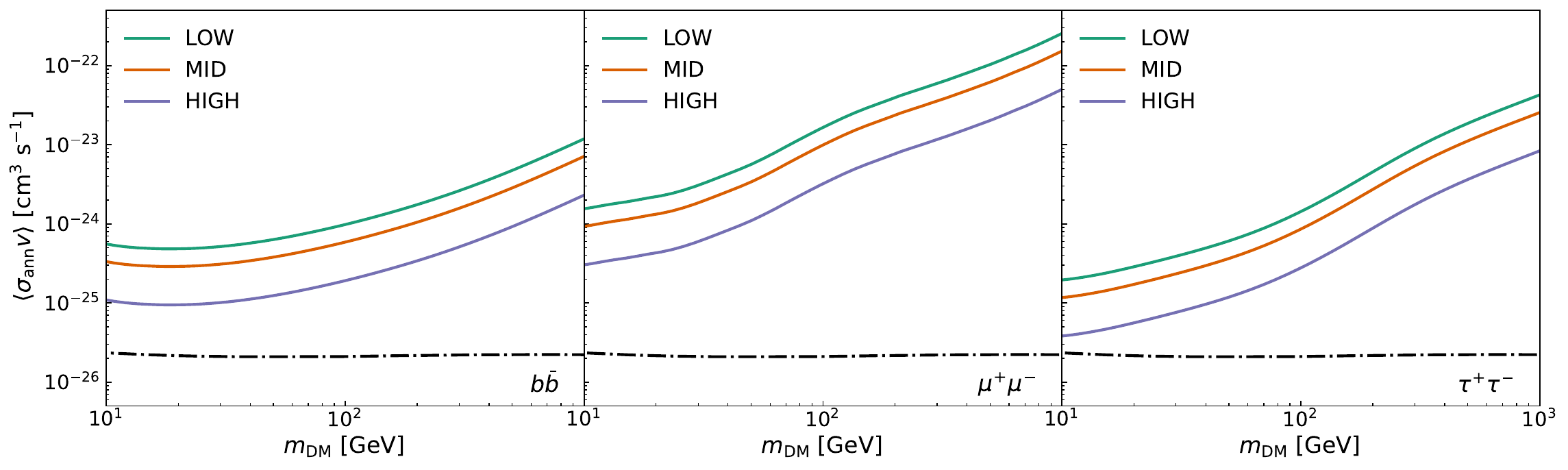}
     \caption{The 95\% upper bounds on the DM annihilation cross-section $\left<\sigma_{\rm ann} v\right>$ for $b\bar{b},~ \mu^-\mu^+$, and $\tau^-\tau^+$ states making use of six redshift bins and ten energy bins. The unresolved blazar contribution to the UGRB is included as the astrophysical background component in the analysis. The thermal relic cross-section for WIMPs \citep{steigman2012precise} is shown as dashed-dot lines. The upper bounds are presented for three DM clumping scenarios: \textsc{low} (green), \textsc{mid} (orange) and \textsc{high} (purple). } 
         \label{Fig:sigmav}
   \end{figure*}

\begin{figure}
  \centering
   \includegraphics[width=\hsize]{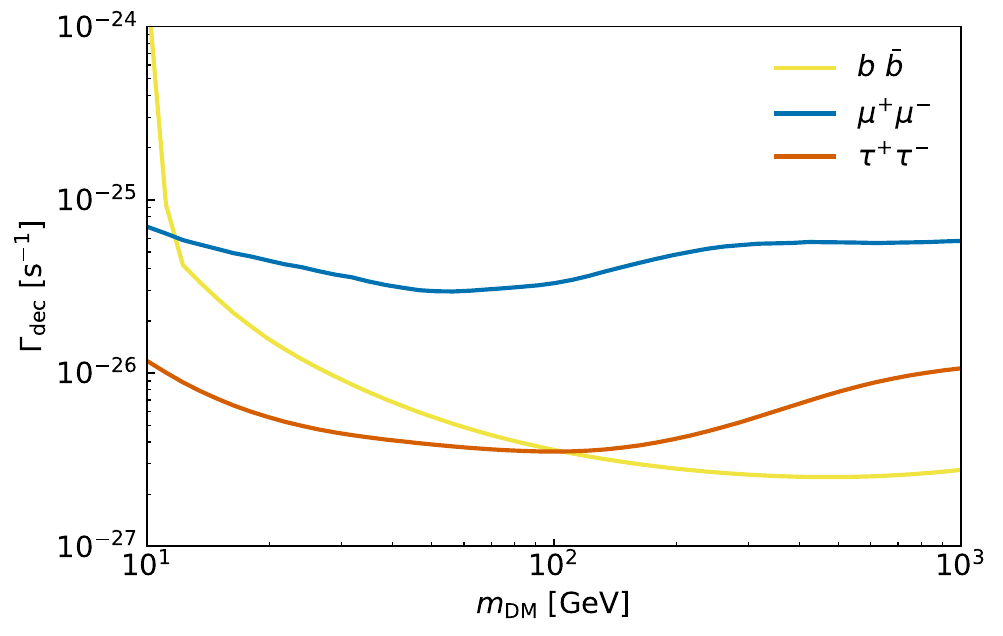}
     \caption{The 95\% upper bounds on the DM decay rate $\Gamma_{\rm dec}$ for $b\bar{b},~ \mu^-\mu^+$, and $\tau^-\tau^+$ states include the unresolved blazar contribution to the UGRB as the astrophysical background.} %\textcolor{red}{fontsize; xticks; Units}}
         \label{Fig:decayrate}
   \end{figure}

% Compare with other results (maybe plot in the same plot) - with Troster 2017 & Paopiamsap 2023
We present our results on the 95\% upper bounds of $\left<\sigma_{\rm ann} v\right>$ and $\Gamma_{\rm dec}$ with $m_{\rm DM}$ in the range of $[10, 1000]$ GeV in Fig. \ref{Fig:sigmav} and Fig. \ref{Fig:decayrate}, respectively. 
Compared with the analysis of \citet{paopiamsap2023constraints}, our constraints are less stringent. Their study focuses on the cross-correlation
between the UGRB and the galaxy overdensity at low redshift
($z \lesssim 0.4$), using wide-area galaxy surveys covering a large
fraction of the extragalactic sky, which yields a high
S/N detection at the $8$–$10\,\sigma$ level. 
In contrast, our study relies on shear measurements over a
smaller effective coverage.

Our results are simplified as we do not account for gamma-ray emission from inverse Compton scattering of CMB photons. 
Compared to previous cross-correlation analyses based on CFHTLenS, RCSLenS, and KiDS data \citep{troster2017cross}, our constraints are generally tighter for \textsc{low} and \textsc{mid} clumping scenarios, while differences at \textsc{high} clumping are likely driven by modelling choices such as substructure prescriptions and the treatment of additional emission components. 
For heavy DM candidates, inverse Compton scattering can significantly contribute to the gamma-ray signal at high energies \citep{ando2016constraining}.
In addition, we adopt a simplified model of the astrophysical background. Incorporating more complete modelling of astrophysical sources and additional emission processes is therefore expected to yield more accurate and potentially tighter constraints in future analyses.

\begin{figure*}
  \centering
 \includegraphics[width=\hsize]{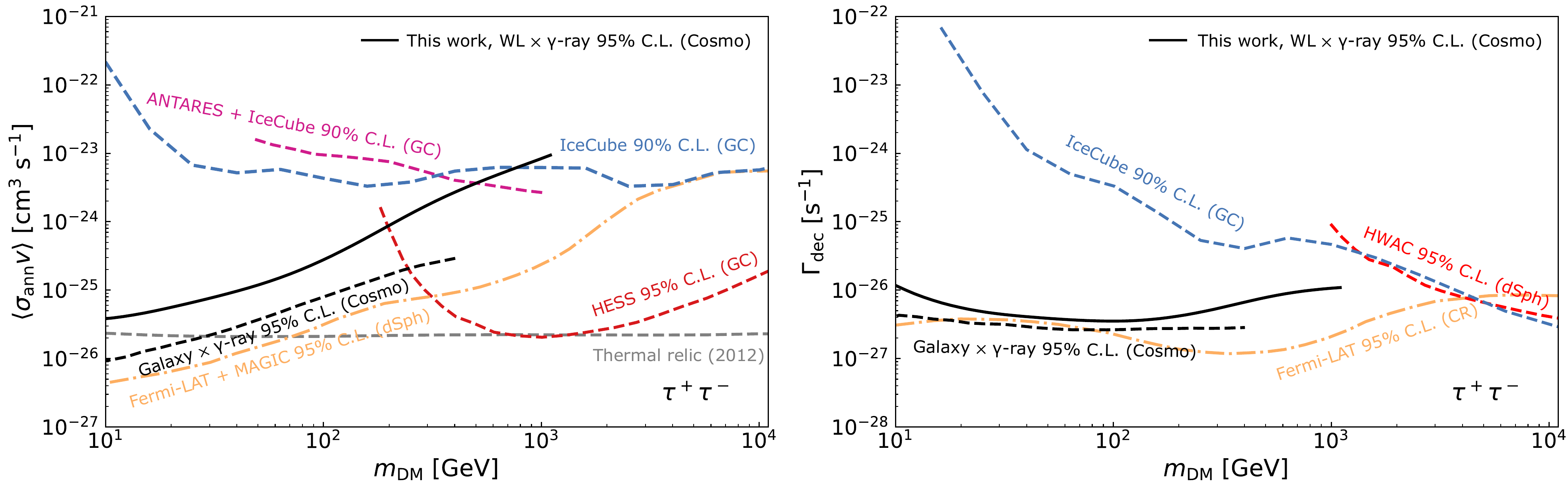}
     \caption{\textit{Left}: Upper bounds on the DM thermally averaged cross-section $\left<\sigma_{\rm ann} v\right>$ of \textsc{high} clumping case for the $\tau^+\tau^-$ state, compared to constraints from local probes. This includes results from neutrino detectors such as IceCube \citep{abbasi2023search} and combined results from ANTARES and IceCube \citep{albert2020combined}; the gamma-ray telescope H.E.S.S. \citep{abdallah2016search}, combined results from Fermi-LAT and MAGIC \citep{ahnen2016limits}; and the red dotted lines represent thermal relic cross-section. \textit{Right}: Upper bounds on the DM decay rate $\Gamma_{\rm dec}$ for the $\tau^+\tau^-$ state, compared to limits from IceCube \citep{aartsen2018search} for neutrino surveys, HAWC \citep{albert2018dark} and searches of cosmic rays and the interstellar medium with Fermi-LAT \citep{ackermann2012fermi} from gamma-ray telescopes. The cosmological constraints from this work (black solid lines) and work from \citet{paopiamsap2023constraints} (black dashed lines) are also shown. GC: Galactic centre; dSph: dwarf spheroidal galaxies; Cosmo: cosmological probe.}
         \label{Fig:comp}
   \end{figure*}

We also compared the 95\% upper bounds of $\left<\sigma_{\rm ann} v\right>$ and $\Gamma_{\rm dec}$ from large-scale cosmological cross-correlation \citep[for the $\tau^+\tau^-$ state from this work and][]{paopiamsap2023constraints}
with constraints derived from local structures as shown in Fig. \ref{Fig:comp}. 
Local probes such as IceCube (neutrinos), H.E.S.S., and MAGIC (gamma-rays) focus on regions of high dark-matter density, such as the Galactic Centre and dwarf spheroidal galaxies, where annihilation or decay signals are expected to be largest. 
By contrast, cosmological analyses constrain dark matter by averaging over a much larger volume, and are less tied to assumptions about individual targets and substructure. As shown in Fig. \ref{Fig:comp}, cosmological methods provide competitive constraints on low $m_{\rm DM}$ for both $\left<\sigma_{\rm ann} v\right>$ and $\Gamma_{\rm dec}$.

\subsection{Forecast of DM parameters with \textit{Euclid}} %% >>> progressing
\label{subsec: forecast} 

Future surveys aiming to measure cosmic shear, covering a wider area and more extensive depth, such as LSST and \textit{Euclid}, are expected to offer improved sensitivity to DM signals. 
In this section, we present a forecast of constraints on the DM annihilation cross-section and decay rate using a \textit{Euclid}-like survey. 
The \textit{Euclid} survey observes about $15~000 \rm ~deg^{2}$ of the extragalactic sky. For this analysis, we adopted ten tomographic bins, together with the survey area, redshift distributions, intrinsic ellipticity dispersion, and expected source number densities specified in table 4 of \citet{blanchard2020euclid}. 

We employed a Fisher matrix formalism to forecast the uncertainties of DM parameters $\left<\sigma_{\rm ann} v\right>$ and $\Gamma_{\rm dec}$ as a function of the logarithmically spaced DM mass $m_{\rm DM}$. The Fisher matrix with Gaussian likelihood is defined as:
\begin{align}
\label{Eq: fisher}
\mathbf{F}_{ij} = \left<\frac{\partial^2 L(\boldsymbol{d}~|~\boldsymbol{p})}{\partial p_i \partial p_j}\right> = \sum_{\ell,\ell'} \frac{\partial \boldsymbol\mu_\ell(\boldsymbol{p})}{\partial p_i} \mathbf{C}^{-1}_{\ell\ell'} \frac{\partial \boldsymbol\mu_{\ell'}(\boldsymbol{p})}{\partial p_j},
\end{align}
where $L(\boldsymbol{d}~|~\boldsymbol{p})$ is the likelihood of observing the data $\boldsymbol{d}$ given the model parameters $\boldsymbol{p}$, $\mathbf{C}_{\ell\ell'}$ is the covariance, and $\boldsymbol\mu_{\ell}(\boldsymbol{p})$ is the theoretical prediction for the cross power spectrum, which represents the stacked vector containing all tomographic–energy combinations at $\ell$.
The marginalised uncertainties on the parameters $p_i$ are given by:
\begin{align}
\label{Eq: estimation}
\sigma(p_i) = \sqrt{(\boldsymbol{\mathrm{F}}^{-1})_{ii}}.
\end{align}

\begin{figure*} 
  \centering
 \includegraphics[width=\hsize]{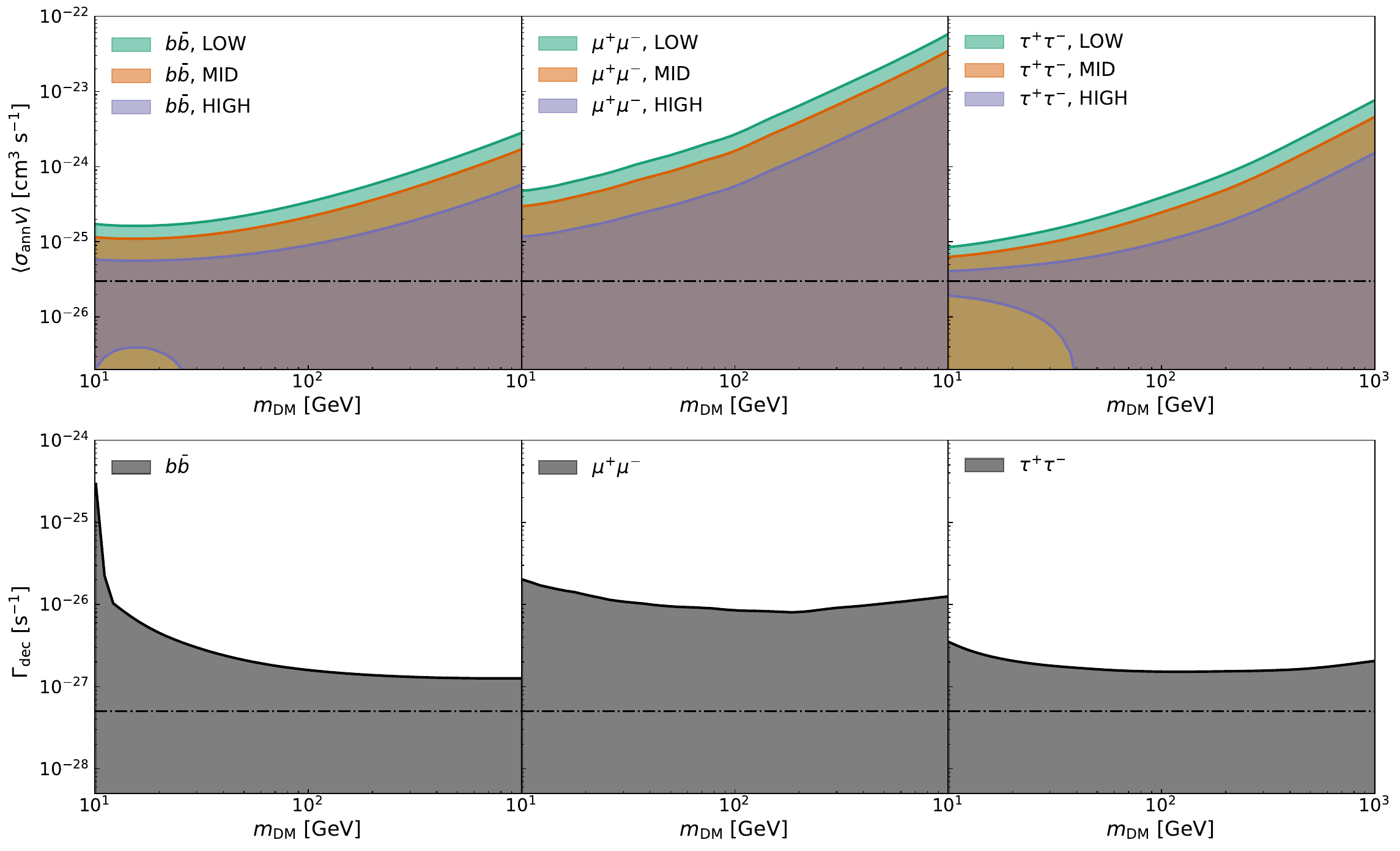}
     \caption{\textit{Upper}: the forecast of 95\% upper and lower bounds on the DM annihilation cross-section $\left<\sigma_{\rm ann} v\right>$ for $b\bar{b},~ \mu^-\mu^+$, and $\tau^-\tau^+$ states from cross-correlation of ten energy bins from Fermi-LAT and ten redshift bins from \textit{Euclid}-like survey. The expected value $\left<\sigma_{\rm ann} v\right> = 3\times 10^{-26}\rm ~cm^3 s^{-1}$ is shown as dot-dashed lines for all cases. \textit{Lower}: the forecast of 95\% upper and lower bounds on the DM decay rate $\Gamma_{\rm dec}$ for $b\bar{b},~ \mu^-\mu^+$, and $\tau^-\tau^+$ states. The dot-dashed lines represent the expected value for the forecast with $\Gamma_{\rm dec} = 5\times10^{-28} \rm ~s^{-1}$.}
         \label{Fig:fisher}
   \end{figure*}

We used the same Fermi–LAT gamma-ray intensity maps as in the measurements. The covariance was calculated using the \texttt{NaMaster}-based analytical estimation detailed in Sect.~\ref{Sec:cov}, and we compared the forecasts using different covariance prescriptions in Appendix~\ref{Sec:forecast_cov}, given that we employed the combined covariance for the measurements. 
The redshift distributions for ten equally populated tomographic bins with edges \{0.001, 0.418, 0.560, 0.678, 0.789, 0.900, 1.019, 1.155, 1.324, 1.576, 2.50\} were modelled as in sect.~3.3.1 in \citet{blanchard2020euclid}. 
We roughly estimated the effective number density for each redshift bin to be the galaxy number density divided by the number of redshift bins, $n_{\rm gal} / N_{z}$, which was $3~\rm arcmin^{-2}$ per bin.
The shape noise was set to $\sigma_\epsilon = 0.3$, the total intrinsic ellipticity dispersion for \textit{Euclid}, where $\sigma_\epsilon^2 = 2\sigma_e^2$ relates it to the per-component ellipticity dispersion.  We adopted the same multipole range as in the KiDS-Legacy measurement ($\ell_{\max}=1500$), which is considered as the pessimistic choice of $\ell_{\max}$ \citep{blanchard2020euclid}.
The Fisher forecast for the cross-correlation between gamma-rays from Fermi-LAT and weak lensing data from \textit{Euclid}-like survey is shown in Fig. \ref{Fig:fisher}, presenting the 95\% bounds of $\left<\sigma_{\rm ann} v\right>$ and $\Gamma_{\rm dec}$ for three different final state particles, where we assume a fiducial DM signal corresponding to $\left<\sigma_{\rm ann} v\right> = 3 \times 10^{-26}~\rm cm^3s^{-1}$ for annihilation and $\Gamma_{\rm dec} = 5\times10^{-28} \rm~s^{-1}$ in the decay case. 
The exclusion of the thermal relic cross-section in the $b\bar{b}$ and $\tau^+\tau^-$ \textsc{high} scenario should not be interpreted as a detection, but rather as a result of the small forecasted covariance in this case. Since these limits are derived within a Fisher forecast framework, they only reflect the statistical sensitivity of the experiment under the assumed model.
Relative to KiDS-Legacy, the \textit{Euclid}-like forecasts tighten the limits by factor of 2 for both $\langle\sigma_{\rm ann} v\rangle$ and $\Gamma_{\rm dec}$ across all final states. This improvement is smaller than expected given the substantial increase in survey area. This is because the sensitivity of the cross-correlation does not scale solely with the weak-lensing data, but is also limited by the properties of the gamma-ray maps. At low energies, the large Fermi–LAT PSF requires larger point-source masking, which reduce the effective sky overlap, while at high energies the limited photon counts lead to large statistical uncertainties. As a result, the improvement in \textit{Euclid} coverage does not propagate into a proportional tightening of the final constraints. 
Within this framework, surveys such as LSST, which achieve comparable effective source number densities and sky coverage, are expected to yield constraints of a similar order.

\section{Conclusions and discussions}
\label{Sec:Conclusions}

In this study, we performed a cross-correlation analysis between the gamma-ray intensity map from 15 years of Fermi-LAT data and weak lensing data from KiDS-Legacy, covering 1347 deg$^{2}$ of the sky. Our aim was to constrain the DM annihilation cross-section $\left<\sigma_{\rm ann} v\right>$ and the decay rate $\Gamma_{\rm dec}$ for WIMP masses between 10 GeV and 1 TeV for three different final states: $b\bar{b}$, $\mu^+\mu^-$, and $\tau^+\tau^-$. We measured the cross-correlation angular power spectrum using six tomographic redshift bins for the weak lensing data and ten energy bins for the gamma-ray data, with an energy range from 0.5 to 1000 GeV. We found no significant detection in the multipole range $200\leq \ell < 1500$.

We modelled the unresolved astrophysical background in the UGRB with a blazar-dominated component, in order to isolate any potential DM contribution. Our analysis demonstrates the 95\% upper bounds for $\left<\sigma_{\rm ann} v\right>$ and $\Gamma_{\rm dec}$. Compared to previous findings, our constraints exhibit greater stringency for the \textsc{low} and \textsc{mid} clumping scenarios. In comparison to the upper bounds from local structure observations, such as neutrino searches by IceCube and gamma-ray observations by H.E.S.S., MAGIC and Fermi-LAT, the cosmological cross-correlation approach is competitive and provides complementary, large-scale constraints on DM properties, particularly in the low-mass region.

A notable aspect of our study is that we do not find a statistically significant UGRB–lensing cross-correlation in the KiDS-Legacy $\times$ Fermi-LAT analysis. 
By contrast, using galaxy clustering as the large-scale structure tracer, \citet{paopiamsap2023constraints} report an $\sim$8-10$\sigma$ detection when cross-correlating with the UGRB. As a robustness check, our pipeline recovers a comparable signal ($\sim$8$\sigma$), consistent with the higher S/N of density tracers relative to shear. 
For weak lensing, DES Y1 found a $5.3\sigma$ detection \citep{ammazzalorso2020detection}, and DES Y3 found an $8.9\sigma$ detection that cross-correlates with Fermi-LAT \citep{thakore2025high}.  
We recovered a DES Y3 UGRB-shear signal at $\sim$6$\sigma$. It corresponds to a null-hypothesis $\chi^2$ test using a jackknife-based covariance within our pipeline. In addition, our analysis uses \texttt{ULTRACLEANVETO} Fermi-LAT events, while DES Y3 adopted \texttt{CLEANVETO} events, which can further affect the quoted significance. Applying the methodology to KiDS-Legacy weak lensing and 15-year Fermi-LAT UGRB yields no significant detection.
This difference can be explained by the effective sky area and the sensitivity to low-$\ell$ modes after gamma-ray processing and masking. Our stricter LAT selections might reduce photon counts and sky fraction and tend to down-weight the lowest multipoles, where DES Y3 derived much of its significance.
Applying the methodology to KiDS-Legacy weak lensing and 15-year Fermi-LAT UGRB yields no significant detection. We also tested alternative multipole binning, and our conclusions were not changed. A summary of the additional validation tests is provided in Appendix~\ref{Sec:compare}.
The more fragmented geometry of the KiDS–Fermi-LAT overlap may increase mask-induced mode coupling. This reduces the number of truly independent low-\(\ell\) modes and, therefore, weakens our leverage on large-scale correlations. 
Differences in tomography and noise properties further modify the effective kernel overlap between lensing and the gamma-ray intensity field. Variations in the redshift distributions, the effective source number density, and shape noise may affect the expected cross-correlation amplitude and uncertainty.
In addition, choices in estimators and modelling, such as the treatment of the Fermi-LAT beam function and the level of diffuse-foreground masking, set the effective weight assigned to different angular scales. Those settings that down weight the lowest multipoles would reduce sensitivity to large-scale correlations. 
Another contributing factor is that the global $\chi^2$ values are slightly lower than expected. As discussed in Sect.~\ref{subsec: results_cross-corr}, this likely reflects a conservative covariance estimation arising from the spatially varying photon noise in the Fermi intensity maps, which should be accounted for in future analyses with higher signal-to-noise data.

A cross-correlation detection between the UGRB and large-scale structure tracers does not necessarily imply a DM detection. Without sufficiently complete modelling of unresolved astrophysical emission and residual large-scale foregrounds, the measured correlation can be explained by conventional source populations. In this context, DES Y3 $\times$ Fermi \citep{thakore2025high} interprets their detection mainly as information on the unresolved blazar contribution. Likewise, \citet{paopiamsap2023constraints} models the astrophysical component using UGRB-galaxy cross-correlations and reports DM constraints as upper limits, obtained by conservatively attributing all the measured signal to a possible DM contribution. 
Although galaxy clustering currently yields a generally higher S/N in UGRB cross-correlation measurements, weak lensing provides a complementary probe. Unlike galaxy tracers, lensing is sensitive to the total matter distribution and does not depend on assumptions about galaxy bias. This makes UGRB–lensing cross-correlations less sensitive to modelling uncertainties in the tracer population, though the statistical uncertainties are larger at present. 

We also forecast the potential improvements that future surveys can bring to DM research. Future surveys, such as the LSST and \textit{Euclid}, with larger coverage and better photometry, will provide higher S/N and more precise  cross-correlation measurements. For a \textit{Euclid}-like configuration, our forecasts indicate about $\sim$2 times tighter constraints on both $\left<\sigma_{\rm ann} v\right>$ and $\Gamma_{\rm dec}$ relative to KiDS-Legacy. On the gamma-ray side, the ongoing observations of Fermi-LAT will increase the total exposure time and reduce photon shot noise, which in turn increases the S/N of the cross-correlation and tightens the DM constraints.

In our analysis, we adopted fixed spectra and modelled the UGRB with blazars only, estimating limits via a maximum-likelihood approach without marginalising over astrophysical-background parameters. 
Given the low S/N in our measurements and the relatively simple parameter space of the DM model from our assumption, this approach simplified the analysis while maintaining computational efficiency. Although this method does not fully account for potential uncertainties in the blazar power spectra, it is sufficient within the scope of our assumptions, as the coupling between the blazar contribution and the DM-related parameters is negligible. Furthermore, focusing on blazars also constitutes a conservative modelling choice for the astrophysical background, avoiding overfitting in the absence of a significant detection.

Future studies, particularly those with higher sensitivity or more significant signals, would benefit from a more detailed treatment of astrophysical contributions. This involves accounting for additional sources of astrophysical background, such as star-forming galaxies, mAGN, and millisecond pulsars, or adopting Bayesian-based techniques and marginalising over the parameters related to astrophysical contributions. These advancements will be critical for refining constraints and deepening our understanding of DM properties.

\begin{acknowledgements}
We thank Bhashin Thakore, David Alonso, Stefano Camera, Marco Regis, Henk Hoekstra, Stefan Fr\"{o}se, Paul-Simon Blomenkamp and Tristan Gradetzke for helpful discussions during the preparation of this manuscript.
SZ acknowledges the support from the Deutsche Forschungsgemeinschaft (DFG) SFB1491. 
H. Hildebrandt is supported by a DFG Heisenberg grant (Hi 1495/5-1), the DFG Collaborative Research Center SFB1491, an ERC Consolidator Grant (No. 770935), and the DLR project 50QE2305. 
ZY, CH and BS acknowledges support from the Max Planck Society and the Alexander von Humboldt Foundation in the framework of the Max Planck-Humboldt Research Award endowed by the Federal Ministry of Education and Research. 
TT acknowledges funding from the Swiss National Science Foundation under the Ambizione project PZ00P2\_193352. 
%AA
MA acknowledges the UK Science and Technology Facilities Council (STFC) under grant number ST/Y002652/1 and the Royal Society under grant numbers RGSR2222268 and ICAR1231094. 
DJB was supported by the Simons Collaboration on ``Learning the Universe’’ and support was provided by Schmidt Sciences, LLC.
MB is supported by the Polish National Science Center through grant no. 2020/38/E/ST9/00395.
D.E. acknowledges the support from DFG Sonderforschungsbereich 1491 Project F5 and BMBF ErUM-Pro.
CH acknowledges the UK Science and Technology Facilities Council (STFC) under grant ST/V000594/1.
BJ acknowledges support by the ERC-selected UKRI Frontier Research Grant EP/Y03015X/1 and by STFC Consolidated Grant ST/V000780/1.
LM acknowledges the financial contribution from the grant PRIN-MUR 2022 20227RNLY3 “The concordance cosmological model: stress-tests with galaxy clusters” supported by Next Generation EU and from the grant ASI n. 2024-10-HH.0 “Attività scientifiche per la missione Euclid – fase E”.
DN acknowledges funding from the European Research Council (ERC) under the European Union’s Horizon 2020 research and innovation program (Grant agreement No. 101053992).
AP acknowledges the support from ``la Caixa” Foundation (ID 100010434, code LCF/BQ/DI24/12070011)
Funding for this work was partially provided by the “Center of Excellence Maria de Maeztu" award to the ICCUB CEX2024-001451-M funded by MICIU/AEI/10.13039/501100011033.
RR is partially supported by an ERC Consolidator Grant (No. 770935).

\end{acknowledgements}

  \bibliographystyle{aa.bst}
  \bibliography{aa.bib}
  
\begin{appendix} 

\onecolumn

\section{Visualisation of the covariance}
\label{App:CovMatrix}

\begin{figure}[H]
\includegraphics[width=\textwidth]{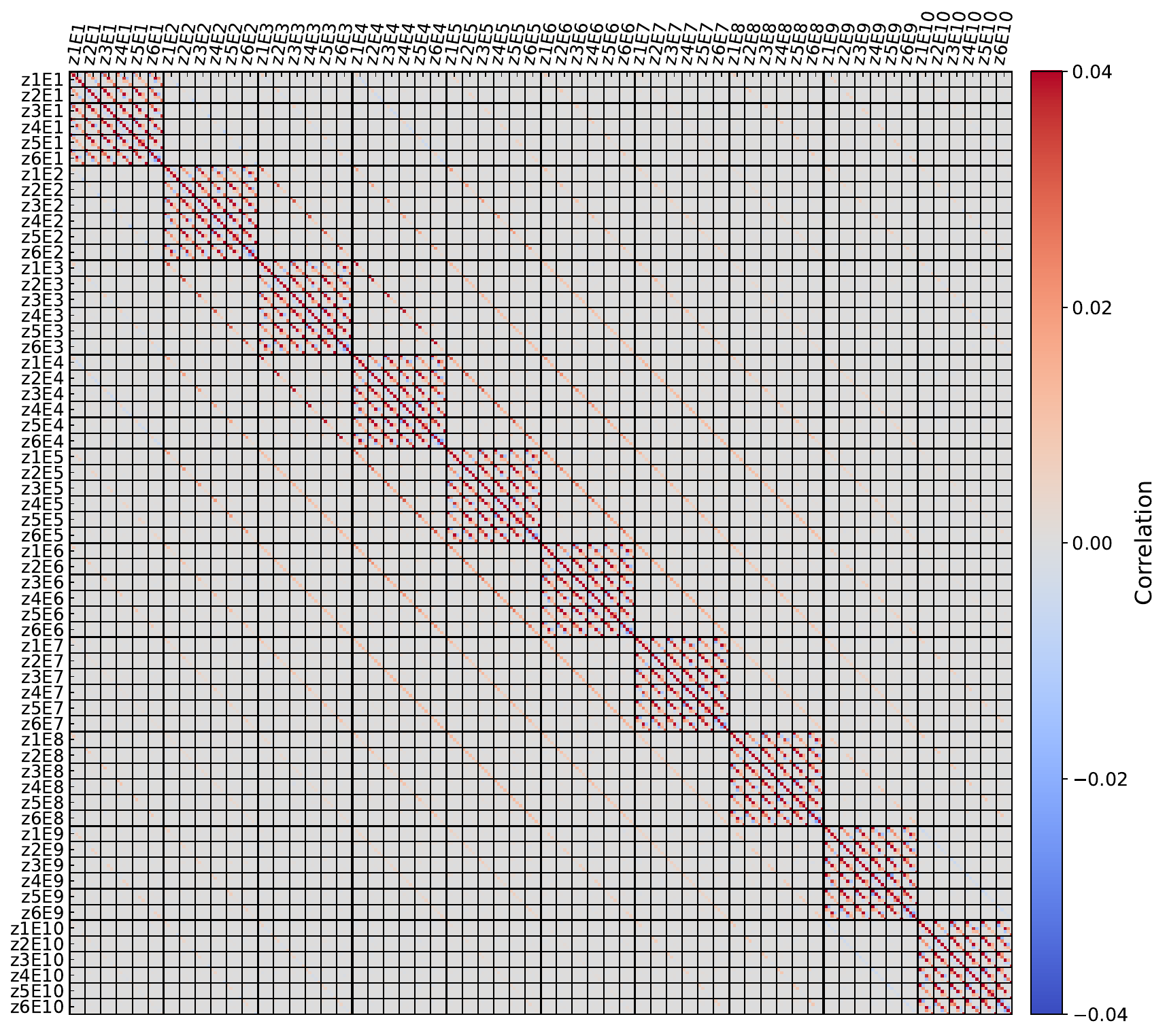}
\caption{Correlation matrix for the cross-correlation between weak lensing data and gamma-ray intensity maps across different redshift and energy bins. The redshift bins (z1 – z6) are the six tomographic bins listed in Table~\ref{Tab:shape_noise}. The energy bins (E1 – E10) span 0.5–1000 GeV from low to high energy.}
\label{Fig:CovMatrix}
\end{figure}

In this appendix, we present a visualisation of the covariance structure used in the analysis.
As described in Sect.~\ref{Sec:cov}, the fiducial covariance is constructed by combining the jackknife variances on the diagonal with the correlation structure estimated from the \texttt{NaMaster} Gaussian covariance.
To illustrate the resulting correlations between multipole, redshift, and energy bins, we show the corresponding correlation matrix in Fig.~\ref{Fig:CovMatrix}.

\section{Forecast from different covariance} 
\label{Sec:forecast_cov}

We adopted a Fisher forecast for a \textit{Euclid}-like survey with wider sky coverage and better photometry to estimate the 95\% upper bounds of the DM parameters, as detailed in Sect.~\ref{subsec: forecast}. 
In the forecasts, we used the analytical covariance detailed in Sect.~\ref{Sec:cov} for the \textit{Euclid}-like survey, whereas we used combined covariance in the measurement. In this section, we compared the forecasts for KiDS-like survey using different covariance estimations and a \textit{Euclid}-like survey with analytical covariance to demonstrate the feasibility and potential bias of using analytical covariance in the Fisher forecast for a \textit{Euclid}-like survey.

\begin{figure*}
  \centering
 \includegraphics[width=\hsize]{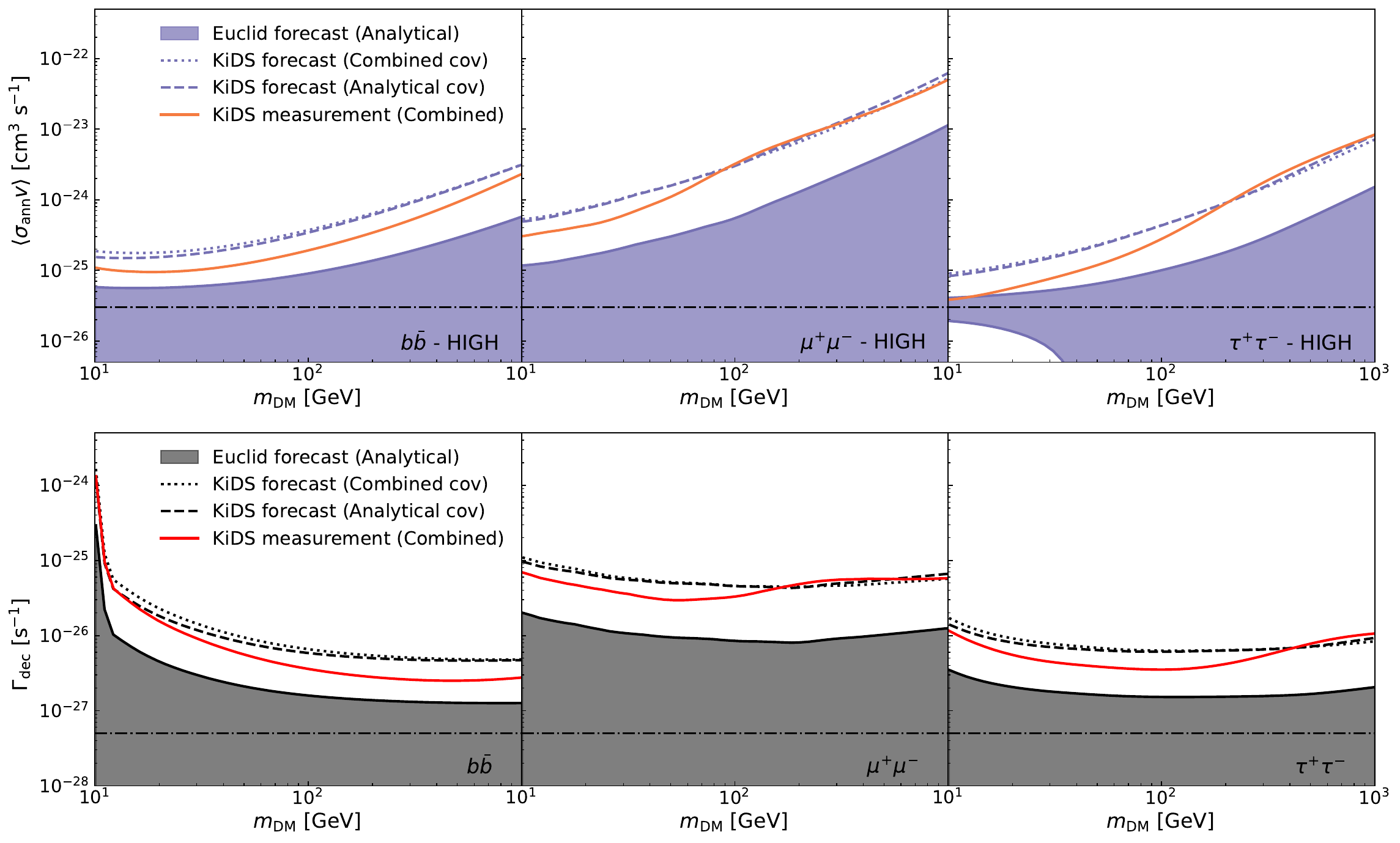}
     \caption{\textit{Upper}: comparison of the forecasts from \textit{Euclid}-like survey with analytical covariance, KiDS-Legacy with analytical and combined covariance and the KiDS-Legacy measurement of 95\% bounds on the DM annihilation cross-section $\left<\sigma_{\rm ann} v\right>$ for three final states of \textsc{high} clumping scenario from combined covariance. The dot-dashed lines represent the expected cross-section $\left<\sigma_{\rm ann} v\right> = 3\times 10^{-26}\rm ~cm^3 s^{-1}$.  \textit{Lower}: forecast for three cases and KiDS-Legacy measurement of 95\% upper and lower bounds on the DM decay rate $\Gamma_{\rm dec}$ for three final states. The model value of decay rate is taken as $\Gamma_{\rm dec} = 5\times10^{-28}\rm ~s^{-1}$. }
         \label{Fig:fisher_comp}
   \end{figure*}

The comparison of the \textsc{high} clumping case of the forecasts $\left<\sigma_{\rm ann} v\right>$ and $\Gamma_{\rm dec}$ for KiDS-like (with combined and analytical covariance) and \textit{Euclid}-like (with analytical covariance), together with the measurements in this study, is shown in Fig. \ref{Fig:fisher_comp}. We used the same combined covariance estimated from the cross angular power spectrum of the Fermi-LAT gamma-rays and KiDS-Legacy weak lensing data as detailed in Sect.~\ref{Sec:cov}.

The upper panels compare 95\% confidence upper bounds on $\left<\sigma_{\rm ann} v\right>$ for three final states in the \textsc{high} clumping case. For KiDS-like surveys, the forecasts (dashed curves: analytical covariance; dotted curves: combined covariance) and the measurements (solid curves: combined covariance) are broadly consistent, with the measurement curves yielding tighter limits than the forecasts in the low $m_{\rm DM}$ region. 
This difference is compatible with statistical fluctuations in the measured cross-correlation with respect to the fiducial model, given that the Fisher forecasts describe the average sensitivity over realisations.
The \textit{Euclid}-like forecasts are uniformly tighter across all channels, as expected from the larger coverage and depth.
Similarly, for the results of $\Gamma_{\rm dec}$ shown in the lower panel, the analytical and combined forecasts are consistent, whereas the measurement shows a larger deviation at low masses for $\mu^+\mu^-$ and $\tau^+\tau^-$ channels, but converges towards the forecasted bounds at higher $m_{\rm DM}$; for $b\bar{b}$ pairs, the measurement becomes tighter than the forecasts at the high-mass end. 
At low $m_{\rm DM}$, the $b\bar{b}$ channel lies close to its kinematic thresholds, so forecasts near the threshold are less robust, which accounts for the small data and forecast differences observed in that regime.

\section{Tests of the discrepancy between DES Y3 and KiDS results} 
\label{Sec:compare}

We performed a series of consistency tests to investigate the origin of the discrepancy between the highly significant cross-correlation signal reported for DES Y3 and the null detection found in this work using KiDS-Legacy shear data. Below, we briefly summarise these tests and the possible causes suggested by their outcomes.

Firstly, the difference between the DES and KiDS results cannot be explained by survey area or angular scale coverage alone. DES Y1, which covers an area comparable to KiDS-Legacy ($\sim$1500 deg$^2$), has already reported a significant detection, indicating that the signal is not driven solely by the larger DES Y3 footprint. Moreover, the DES analysis probes a wide range of angular scales using the tangential shear statistic, with a significant signal detected from arcminute to degree scales. These scales overlap substantially with those probed in our pseudo-$C_\ell$ analysis, suggesting that the discrepancy is not due to DES accessing scales unavailable in KiDS.

Secondly, we tested whether differences in the construction and processing of the Fermi–LAT maps could explain the discrepancy. Using the UGRB maps and pipeline choices from \citet{paopiamsap2023constraints}, we reproduced their galaxy–gamma-ray cross-correlation signal when correlating the \citet{paopiamsap2023constraints} UGRB maps with their galaxy samples (2MPZ and WISE–SuperCOSMOS). Repeating the same cross-correlation using our fiducial Fermi–LAT maps yields a consistent signal level, with small differences attributable to the different energy binning adopted in this work. We further applied the same Fermi UGRB maps to weak-lensing shear tracers. When cross-correlated with DES shear, we recover a significant detection with $\sim$$6 \sigma$ using cross correlations. In contrast, when cross-correlating with KiDS shear, we obtain a null result, and this remains true whether we use the \citet{paopiamsap2023constraints} UGRB maps or our fiducial Fermi–LAT maps. Taken together, these tests indicate that our implementation is able to reproduce the literature galaxy–gamma-ray measurements and the DES shear–gamma-ray detection when using the corresponding external data products. Moreover, the KiDS shear null detection is not driven by the particular choice of the Fermi–LAT map but persists across independent gamma-ray map constructions.

Finally, differences between DES and KiDS shear measurements are unlikely to be the dominant cause. The consistency between DES Y3 and KiDS-1000 cosmic shear results has been extensively tested and discussed in the literature (see, e.g., \citealt{DK2023OJAp....6E..36D}), with no evidence of significant systematic offsets that could account for the observed discrepancy in the gamma-ray cross-correlation.

Based on the differences observed between the KiDS-N and KiDS-S footprints, our current working hypothesis is that foregrounds and large-scale anisotropies in the Fermi–LAT sky, arising from Galactic emission and spatially inhomogeneous exposure, lead to an inhomogeneous cross-correlation signal across the sky.
Under this inhomogeneity interpretation, the effective signal amplitude within the KiDS footprint could significantly differ from DES, as this signal includes more than just the homogeneous signal expected from WIMP annihilation or decay.
This hypothesis will be tested more robustly with upcoming wide-area weak-lensing surveys such as \textit{Euclid} and LSST, which will allow nearly full-sky cross-correlation analyses and a more direct assessment of spatial variations in the signal.

\end{appendix}

\end{document}